\documentclass[10pt,letterpaper]{article}
\usepackage{opex3}
\usepackage{cite}
\usepackage{graphicx}
\usepackage{dcolumn}
\usepackage{bm}
\usepackage{hyperref}
\usepackage{subfigure}
\usepackage{color}
\usepackage{empheq}
\usepackage{lipsum}
\usepackage{amssymb,amsbsy,exscale,amsmath,amsfonts,amssymb,amscd}
\usepackage{stmaryrd}
\usepackage{makeidx}
\usepackage{multirow}

\newcommand{\secref}[1]{$\S$\ref{#1}}

\definecolor{blue}{rgb}{0,0,1}

\newcommand{\figref}[1]{\figurename~\ref{#1}}
\newcommand{\citeasnoun}[1]{Ref.~\citenum{#1}}

\renewcommand{\vec}[1]{\mathbf{#1}}

\begin{document}

\title{Robust topology optimization of three-dimensional photonic-crystal band-gap structures}

\author{H. Men$^{1*}$, K. Y. K. Lee, R. M. Freund$^3$, J. Peraire$^1$, S. G. Johnson$^2$}
\address{$^1$ Department of Aeronautics and Astronautics, Massachusetts Institute of Technology, Cambridge, MA 02139, USA}
\address{$^2$ Department of Mathematics, Massachusetts Institute of Technology, Cambridge, MA 02139, USA}
\address{$^3$ Sloan School of Management, Massachusetts Institute of Technology, Cambridge, MA 02142, USA}
\email{*abbymen@mit.edu} 


\begin{abstract}
We perform full 3D topology optimization (in which ``every voxel'' of the unit cell is a degree of freedom) of photonic-crystal structures in order to find optimal omnidirectional band gaps for various symmetry groups, including fcc (including diamond), bcc, and simple-cubic lattices. Even without imposing the constraints of any fabrication process, the resulting optimal gaps are only slightly larger than previous hand designs, suggesting that current photonic crystals are nearly optimal in this respect.  However, optimization can discover new structures, e.g. a new fcc structure with the same symmetry but slightly larger gap than the well known inverse opal, which may offer new degrees of freedom to future fabrication technologies. Furthermore, our band-gap optimization is an illustration of a computational approach to 3D dispersion engineering which is applicable to many other problems in optics, based on a novel semidefinite-program formulation for nonconvex eigenvalue optimization combined with other techniques such as a simple approach to impose symmetry constraints. We also demonstrate a technique for \emph{robust} topology optimization, in which some uncertainty is included in each voxel and we optimize the worst-case gap, and we show that the resulting band gaps have increased robustness to systematic fabrication errors.
\end{abstract}


\bibliography{3dbg_ref}

\begin{thebibliography}{10}
\newcommand{\enquote}[1]{``#1''}

\bibitem{dobson1999maximizing}
D.~C. Dobson and S.~J. Cox, \enquote{Maximizing band gaps in two-dimensional
  photonic crystals,} SIAM Journal on Applied Mathematics \textbf{59},
  2108--2120 (1999).

\bibitem{cox2000bso}
S.~J. Cox and D.~C. Dobson, \enquote{{Band structure optimization of
  two-dimensional photonic crystals in H-polarization},} Journal of
  Computational Physics \textbf{158}, 214--224 (2000).

\bibitem{shen2002large}
L.~F. Shen, S.~L. He, and S.~H. Xiao, \enquote{Large absolute band gaps in
  two-dimensional photonic crystals formed by large dielectric pixels,}
  Physical Review B \textbf{66}, 165315 (2002).

\bibitem{kao2005mbg}
C.~Y. Kao, S.~J. Osher, and E.~Yablonovitch, \enquote{{Maximizing band gaps in
  two-dimensional photonic crystals by using level set methods},} Applied
  Physics B: Lasers and Optics \textbf{81}, 235--244 (2005).

\bibitem{halkjaer2006maximizing}
S.~Halkj{\ae}r, O.~Sigmund, and J.~S. Jensen, \enquote{Maximizing band gaps in
  plate structures,} Structural and Multidisciplinary Optimization \textbf{32},
  263--275 (2006).

\bibitem{sigmund2008geometric}
O.~Sigmund and K.~Hougaard, \enquote{Geometric properties of optimal photonic
  crystals,} Physical Review Letters \textbf{100}, 153904 (2008).

\bibitem{men2010bandgap}
H.~Men, N.~C. Nguyen, R.~M. Freund, P.~A. Parrilo, and J.~Peraire,
  \enquote{Bandgap optimization of two-dimensional photonic crystals using
  semidefinite programming and subspace methods,} Journal of Computational
  Physics \textbf{229}, 3706--3725 (2010).

\bibitem{maldovan2004diamond}
M.~Maldovan and E.~L. Thomas, \enquote{Diamond-structured photonic crystals,}
  Nature Materials \textbf{3}, 593--600 (2004).

\bibitem{sozuer1992photonic}
H.~S{\"o}z{\"u}er, J.~Haus, and R.~Inguva, \enquote{Photonic bands: Convergence
  problems with the plane-wave method,} Physical Review B \textbf{45}, 13962
  (1992).

\bibitem{joannopoulos2011photonic}
J.~D. Joannopoulos, S.~G. Johnson, J.~N. Winn, and R.~D. Meade, \emph{Photonic
  Crystals: Molding the Flow of Light} (Princeton university press, 2011).

\bibitem{men2014fabrication}
H.~Men, R.~M. Freund, N.~C. Nguyen, J.~Saa-Seoane, and J.~Peraire,
  \enquote{Fabrication-adaptive optimization with an application to photonic
  crystal design,} Operations Research \textbf{62}, 418--434 (2014).

\bibitem{bertsimas2007robust}
D.~Bertsimas, O.~Nohadani, and K.~M. Teo, \enquote{Robust optimization in
  electromagnetic scattering problems,} Journal of Applied Physics
  \textbf{101}, 074507 (2007).

\bibitem{mutapcic2009robust}
A.~Mutapcic, S.~Boyd, A.~Farjadpour, S.~G. Johnson, and Y.~Avniel,
  \enquote{Robust design of slow-light tapers in periodic waveguides,}
  Engineering Optimization \textbf{41}, 365--384 (2009).

\bibitem{sigmund2009manufacturing}
O.~Sigmund, \enquote{Manufacturing tolerant topology optimization,} Acta
  Mechanica Sinica \textbf{25}, 227--239 (2009).

\bibitem{bertsimas2010robust}
D.~Bertsimas, O.~Nohadani, and K.~M. Teo, \enquote{Robust optimization for
  unconstrained simulation-based problems,} Operations Research \textbf{58},
  161--178 (2010).

\bibitem{wang2011robust}
F.~Wang, J.~S. Jensen, and O.~Sigmund, \enquote{Robust topology optimization of
  photonic crystal waveguides with tailored dispersion properties,} JOSA B
  \textbf{28}, 387--397 (2011).

\bibitem{wang2011projection}
F.~Wang, B.~S. Lazarov, and O.~Sigmund, \enquote{On projection methods,
  convergence and robust formulations in topology optimization,} Structural and
  Multidisciplinary Optimization \textbf{43}, 767--784 (2011).

\bibitem{schevenels2011robust}
M.~Schevenels, B.~S. Lazarov, and O.~Sigmund, \enquote{Robust topology
  optimization accounting for spatially varying manufacturing errors,} Computer
  Methods in Applied Mechanics and Engineering \textbf{200}, 3613--3627 (2011).

\bibitem{oskooiMu2012robust}
A.~Oskooi, A.~Mutapcic, S.~Noda, J.~D. Joannopoulos, S.~P. Boyd, and S.~G.
  Johnson, \enquote{Robust optimization of adiabatic tapers for coupling to
  slow-light photonic-crystal waveguides,} Optics Express \textbf{20},
  21558--21575 (2012).

\bibitem{elesin2012design}
Y.~Elesin, B.~S. Lazarov, J.~S. Jensen, and O.~Sigmund, \enquote{Design of
  robust and efficient photonic switches using topology optimization,}
  Photonics and Nanostructures-Fundamentals and Applications \textbf{10},
  153--165 (2012).

\bibitem{fan1994design}
S.~Fan, P.~R. Villeneuve, R.~D. Meade, and J.~D. Joannopoulos, \enquote{Design
  of three-dimensional photonic crystals at submicron lengthscales,} Applied
  physics letters \textbf{65}, 1466--1468 (1994).

\bibitem{doosje2000pbo}
M.~Doosje, B.~J. Hoenders, and J.~Knoester, \enquote{{Photonic bandgap
  optimization in inverted fcc photonic crystals},} Journal of the Optical
  Society of America B \textbf{17}, 600--606 (2000).

\bibitem{biswas2002three}
R.~Biswas, M.~Sigalas, K.~Ho, and S.~Lin, \enquote{Three-dimensional photonic
  band gaps in modified simple cubic lattices,} Physical Review B \textbf{65},
  205121 (2002).

\bibitem{maldovan2002photonic}
M.~Maldovan, A.~Urbas, N.~Yufa, W.~Carter, and E.~Thomas, \enquote{Photonic
  properties of bicontinuous cubic microphases,} Physical Review B \textbf{65},
  165123 (2002).

\bibitem{toader2003photonic}
O.~Toader, M.~Berciu, and S.~John, \enquote{Photonic band gaps based on
  tetragonal lattices of slanted pores,} Physical Review Letters \textbf{90},
  233901 (2003).

\bibitem{maldovan2003exploring}
M.~Maldovan, C.~K. Ullal, W.~C. Carter, and E.~L. Thomas, \enquote{Exploring
  for 3d photonic bandgap structures in the 11 fcc space groups,} Nature
  Materials \textbf{2}, 664--667 (2003).

\bibitem{maldovan2005photonic}
M.~Maldovan and E.~Thomas, \enquote{Photonic crystals: six connected dielectric
  networks with simple cubic symmetry,} JOSA B \textbf{22}, 466--473 (2005).

\bibitem{bendsoe2003topology}
M.~P. Bends{\o}e and O.~Sigmund, \emph{Topology Optimization: Theory, Methods
  and Applications} (Springer, 2003).

\bibitem{burger2004inverse}
M.~Burger, S.~J. Osher, and E.~Yablonovitch, \enquote{Inverse problem
  techniques for the design of photonic crystals,} IEICE Transactions on
  Electronics \textbf{87}, 258--265 (2004).

\bibitem{burger2005survey}
M.~Burger and S.~J. Osher, \enquote{A survey on level set methods for inverse
  problems and optimal design,} European Journal of Applied Mathematics
  \textbf{16}, 263 (2005).

\bibitem{he2007incorporating}
L.~He, C.~Y. Kao, and S.~J. Osher, \enquote{Incorporating topological
  derivatives into shape derivatives based level set methods,} Journal of
  Computational Physics \textbf{225}, 891--909 (2007).

\bibitem{sigmund2003systematic}
O.~Sigmund and J.~S. Jensen, \enquote{Systematic design of phononic band--gap
  materials and structures by topology optimization,} Philosophical
  Transactions of the Royal Society of London. Series A: Mathematical, Physical
  and Engineering Sciences \textbf{361}, 1001--1019 (2003).

\bibitem{jensen2004systematic}
J.~S. Jensen and O.~Sigmund, \enquote{Systematic design of photonic crystal
  structures using topology optimization: Low-loss waveguide bends,} Applied
  Physics Letters \textbf{84}, 2022--2024 (2004).

\bibitem{watanabe2006broadband}
Y.~Watanabe, Y.~Sugimoto, N.~Ikeda, N.~Ozaki, A.~Mizutani, Y.~Takata,
  Y.~Kitagawa, and K.~Asakawa, \enquote{Broadband waveguide intersection with
  low crosstalk in two-dimensional photonic crystal circuits by using topology
  optimization,} Optics express \textbf{14}, 9502--9507 (2006).

\bibitem{liang2013formulation}
X.~Liang and S.~G. Johnson, \enquote{Formulation for scalable optimization of
  microcavities via the frequency-averaged local density of states,} Optics
  Express \textbf{21}, 30812--30841 (2013).

\bibitem{men2010optimal}
H.~Men, \enquote{Optimal design of photonic crystals,} Ph.D. thesis, National
  University of Singapore (2010).

\bibitem{bendsoe1988generating}
M.~P. Bends{\o}e and N.~Kikuchi, \enquote{Generating optimal topologies in
  structural design using a homogenization method,} Computer Methods in Applied
  Mechanics and Engineering \textbf{71}, 197--224 (1988).

\bibitem{bendsoe1995optimization}
M.~P. Bends{\o}e, \emph{Optimization of Structural Topology, Shape, and
  Material} (Springer, Berlin, 1995).

\bibitem{bendsoe1999material}
M.~P. Bends{\o}e and O.~Sigmund, \enquote{Material interpolation schemes in
  topology optimization,} Archive of Applied Mechanics \textbf{69}, 635--654
  (1999).

\bibitem{stolpe2001alternative}
M.~Stolpe and K.~Svanberg, \enquote{An alternative interpolation scheme for
  minimum compliance topology optimization,} Structural and Multidisciplinary
  Optimization \textbf{22}, 116--124 (2001).

\bibitem{bruns2005reevaluation}
T.~Bruns, \enquote{A reevaluation of the simp method with filtering and an
  alternative formulation for solid--void topology optimization,} Structural
  and Multidisciplinary Optimization \textbf{30}, 428--436 (2005).

\bibitem{seyranian1994multiple}
A.~P. Seyranian, E.~Lund, and N.~Olhoff, \enquote{Multiple eigenvalues in
  structural optimization problems,} Structural Optimization \textbf{8},
  207--227 (1994).

\bibitem{cox1995generalized}
S.~J. Cox, \enquote{The generalized gradient at a multiple eigenvalue,} Journal
  of Functional Analysis \textbf{133}, 30--40 (1995).

\bibitem{beyer2007robust}
H.~G. Beyer and B.~Sendhoff, \enquote{Robust optimization--a comprehensive
  survey,} Computer Methods in Applied Mechanics and Engineering \textbf{196},
  3190--3218 (2007).

\bibitem{ben2009robust}
A.~Ben-Tal, L.~El~Ghaoui, and A.~Nemirovski, \emph{Robust Optimization}
  (Princeton University Press, 2009).

\bibitem{bertsimas2011theory}
D.~Bertsimas, D.~B. Brown, and C.~Caramanis, \enquote{Theory and applications
  of robust optimization,} SIAM Review \textbf{53}, 464--501 (2011).

\bibitem{harrison2007occurrence}
J.~Harrison, P.~Kuchment, A.~Sobolev, and B.~Winn, \enquote{On occurrence of
  spectral edges for periodic operators inside the brillouin zone,} Journal of
  Physics A: Mathematical and Theoretical \textbf{40}, 7597 (2007).

\bibitem{rodriguez2005disorder}
A.~Rodriguez, M.~Ibanescu, J.~D. Joannopoulos, and S.~G. Johnson,
  \enquote{Disorder-immune confinement of light in photonic-crystal cavities,}
  Optics Letters \textbf{30}, 3192--3194 (2005).

\bibitem{johnson2001block}
S.~G. Johnson and J.~D. Joannopoulos, \enquote{Block-iterative frequency-domain
  methods for {Maxwell}'s equations in a planewave basis,} Optics Express
  \textbf{8}, 173--190 (2001).

\bibitem{hahn2006international}
T.~Hahn, U.~Shmueli, A.~J.~C. Wilson, and E.~Prince, \emph{International Tables
  for Crystallography}, vol. A, Space-group symmetry (2006).

\bibitem{mosek2013mosek}
Mosek, \enquote{The mosek optimization software,} Online at
  http://www.mosek.com  (2013).

\bibitem{chutinan1998spiral}
A.~Chutinan and S.~Noda, \enquote{Spiral three-dimensional photonic-band-gap
  structure,} Physical Review B \textbf{57}, R2006 (1998).

\bibitem{busch1998photonic}
K.~Busch and S.~John, \enquote{Photonic band gap formation in certain
  self-organizing systems,} Physical Review E \textbf{58}, 3896 (1998).

\bibitem{johnson2000three}
S.~G. Johnson and J.~Joannopoulos, \enquote{Three-dimensionally periodic
  dielectric layered structure with omnidirectional photonic band gap,} Applied
  Physics Letters \textbf{77}, 3490--3492 (2000).

\bibitem{ho1990existence}
K.~Ho, C.~Chan, and C.~Soukoulis, \enquote{Existence of a photonic gap in
  periodic dielectric structures,} Physical Review Letters \textbf{65}, 3152
  (1990).

\bibitem{kosaka1998superprism}
H.~Kosaka, T.~Kawashima, A.~Tomita, M.~Notomi, T.~Tamamura, T.~Sato, and
  S.~Kawakami, \enquote{Superprism phenomena in photonic crystals,} Physical
  Review B \textbf{58}, R10096 (1998).

\bibitem{lin1996highly}
S.-Y. Lin, V.~Hietala, L.~Wang, and E.~Jones, \enquote{Highly dispersive
  photonic band-gap prism,} Optics letters \textbf{21}, 1771--1773 (1996).

\bibitem{kosaka1999superprism}
H.~Kosaka, T.~Kawashima, A.~Tomita, M.~Notomi, T.~Tamamura, T.~Sato, and
  S.~Kawakami, \enquote{Superprism phenomena in photonic crystals: Toward
  microscale lightwave circuits,} Journal of lightwave technology \textbf{17},
  2032 (1999).

\bibitem{wu2002superprism}
L.~Wu, M.~Mazilu, T.~Karle, and T.~F. Krauss, \enquote{Superprism phenomena in
  planar photonic crystals,} Quantum Electronics, IEEE Journal of \textbf{38},
  915--918 (2002).

\bibitem{luo2004superprism}
C.~Luo, M.~Solja{\v{c}}i{\'c}, and J.~Joannopoulos, \enquote{Superprism effect
  based on phase velocities,} Optics letters \textbf{29}, 745--747 (2004).

\bibitem{serbin2005superprism}
J.~Serbin and M.~Gu, \enquote{Superprism phenomena in polymeric woodpile
  structures,} Journal of applied physics \textbf{98}, 123101 (2005).

\bibitem{kosaka1999self}
H.~Kosaka, T.~Kawashima, A.~Tomita, M.~Notomi, T.~Tamamura, T.~Sato, and
  S.~Kawakami, \enquote{Self-collimating phenomena in photonic crystals,}
  Applied Physics Letters \textbf{74}, 1212--1214 (1999).

\bibitem{witzens2002self}
J.~Witzens, M.~Loncar, and A.~Scherer, \enquote{Self-collimation in planar
  photonic crystals,} Selected Topics in Quantum Electronics, IEEE Journal of
  \textbf{8}, 1246--1257 (2002).

\bibitem{wu2003beam}
L.~Wu, M.~Mazilu, and T.~F. Krauss, \enquote{Beam steering in planar-photonic
  crystals: from superprism to supercollimator,} Journal of lightwave
  technology \textbf{21}, 561 (2003).

\bibitem{prather2004dispersion}
D.~W. Prather, S.~Shi, D.~M. Pustai, C.~Chen, S.~Venkataraman, A.~Sharkawy,
  G.~Schneider, and J.~Murakowski, \enquote{Dispersion-based optical routing in
  photonic crystals,} Optics letters \textbf{29}, 50--52 (2004).

\bibitem{shin2005conditions}
J.~Shin and S.~Fan, \enquote{Conditions for self-collimation in
  three-dimensional photonic crystals,} Optics letters \textbf{30}, 2397--2399
  (2005).

\bibitem{lu2006experimental}
Z.~Lu, S.~Shi, J.~A. Murakowski, G.~J. Schneider, C.~A. Schuetz, and D.~W.
  Prather, \enquote{Experimental demonstration of self-collimation inside a
  three-dimensional photonic crystal,} Physical review letters \textbf{96},
  173902 (2006).

\bibitem{rakich2006achieving}
P.~T. Rakich, M.~S. Dahlem, S.~Tandon, M.~Ibanescu, M.~Solja{\v{c}}i{\'c},
  G.~S. Petrich, J.~D. Joannopoulos, L.~A. Kolodziejski, and E.~P. Ippen,
  \enquote{Achieving centimetre-scale supercollimation in a large-area
  two-dimensional photonic crystal,} Nature materials \textbf{5}, 93--96
  (2006).

\bibitem{birks1999dispersion}
T.~Birks, D.~Mogilevtsev, J.~Knight, and P.~S.~J. Russell, \enquote{Dispersion
  compensation using single-material fibers,} Photonics Technology Letters,
  IEEE \textbf{11}, 674--676 (1999).

\bibitem{shen2003design}
L.~P. Shen, W.~P. Huang, G.~X. Chen, and S.~S. Jian, \enquote{Design and
  optimization of photonic crystal fibers for broad-band dispersion
  compensation,} Photonics Technology Letters, IEEE \textbf{15}, 540--542
  (2003).

\bibitem{ni2004dual}
Y.~Ni, L.~Zhang, L.~An, J.~Peng, and C.~Fan, \enquote{Dual-core photonic
  crystal fiber for dispersion compensation,} IEEE photonics technology letters
  \textbf{16}, 1516--1518 (2004).

\bibitem{zsigri2004novel}
B.~Zsigri, J.~L{\ae}gsgaard, and A.~Bjarklev, \enquote{A novel photonic crystal
  fibre design for dispersion compensation,} Journal of Optics A: Pure and
  Applied Optics \textbf{6}, 717 (2004).

\bibitem{fiore1998phase}
A.~Fiore, V.~Berger, E.~Rosencher, P.~Bravetti, and J.~Nagle, \enquote{Phase
  matching using an isotropic nonlinear optical material,} Nature \textbf{391},
  463--466 (1998).

\bibitem{berger1998nonlinear}
V.~Berger, \enquote{Nonlinear photonic crystals,} Physical Review Letters
  \textbf{81}, 4136 (1998).

\bibitem{saltiel2000phase}
S.~Saltiel and Y.~S. Kivshar, \enquote{Phase matching in nonlinear $\chi^{(2)}$
  photonic crystals,} Optics letters \textbf{25}, 1204--1206 (2000).

\bibitem{notomi2000theory}
M.~Notomi, \enquote{Theory of light propagation in strongly modulated photonic
  crystals: Refractionlike behavior in the vicinity of the photonic band gap,}
  Physical Review B \textbf{62}, 10696 (2000).

\bibitem{luo2002pho}
C.~Luo, S.~G. Johnson, and J.~Joannopoulos, \enquote{All-angle negative
  refraction in a three-dimensionally periodic photonic crystal,} Applied
  physics letters \textbf{81}, 2352--2354 (2002).

\bibitem{parimi2003photonic}
P.~V. Parimi, W.~T. Lu, P.~Vodo, and S.~Sridhar, \enquote{Photonic crystals:
  Imaging by flat lens using negative refraction,} Nature \textbf{426},
  404--404 (2003).

\bibitem{prather2006photonic}
D.~W. Prather, S.~Shi, J.~Murakowski, G.~J. Schneider, A.~Sharkawy, C.~Chen,
  and B.~Miao, \enquote{Photonic crystal structures and applications:
  Perspective, overview, and development,} Selected Topics in Quantum
  Electronics, IEEE Journal of \textbf{12}, 1416--1437 (2006).

\end{thebibliography}
\bibliographystyle{osajnl}

\section{Introduction}
\label{sec_intro}
In this paper, we present the first fully three-dimensional (3D) robust topology optimization (in which every voxel is a degree of freedom) of complete photonic band gaps in 3D photonic crystals, in contrast to earlier band-gap topology-optimization work \cite{dobson1999maximizing, cox2000bso, shen2002large, kao2005mbg,halkjaer2006maximizing,sigmund2008geometric,men2010bandgap} that was limited to two dimensions (2D) and did not address robustness to manufacturing defects. Our results in \secref{subsec_opt} confirm that, for diamond symmetry, known ``hand-designed'' 3D crystal structures \cite{maldovan2004diamond}, appear to be close to optimal with respect to the fractional band gap. However, the optimization also appears to discover some previously unknown structures for other symmetry groups, including a new fcc-symmetry structure that has a larger gap for the same bands than the well-known ``inverse-opal'' design~\cite{sozuer1992photonic} (although its gap is still smaller than diamond).   On the one hand, cases where the optimization yields structures that are reminiscent of previous hand designs with only slightly larger gaps, despite searching a large space of continuously varying structures without regard for ease of fabrication, suggest limitations on the prospects for future improvements in band-gap sizes with conventional dielectric materials, in which we find gaps for index contrasts $\ge 1.9:1$ (in \secref{subsec_idx}) similar to previous authors~\cite{joannopoulos2011photonic}.  On the other hand, the ability of optimization to discover new designs for a given symmetry group may inspire exploration of new fabrication technologies that are better suited to those topologies (or a distorted version thereof) than to previous structures.  Moreover, our optimization approach (in \secref{sec_methods}) builds on and illustrates our recent subspace and semidefinite-program (SDP) formulation \cite{men2010bandgap} that is applicable to many other dispersion-relation design problems. Our approach also demonstrates our new ``fabrication adaptivity'' (FA) algorithm \cite{men2014fabrication} for \emph{robust} topology optimization (see \secref{subsec_fa})---optimization of the ``worst case'' subject to manufacturing and other uncertainties---which differs from previous work on robust optimization in electromagnetism \cite{bertsimas2007robust, mutapcic2009robust, sigmund2009manufacturing, bertsimas2010robust, wang2011robust, wang2011projection, schevenels2011robust, oskooiMu2012robust, elesin2012design} by taking into account the material constraints that arise in topology optimization. We also describe a new technique for imposing complex symmetry constraints (e.g. diamond symmetry) which retains the simplicity of an orthogonal grid of degrees of freedom (in \secref{sec_methods}), and allows us to explore a much wider variety of symmetry groups than were considered in most previous topology-optimization work.

\begin{figure}[h]
\centering
\includegraphics[scale = 0.35]{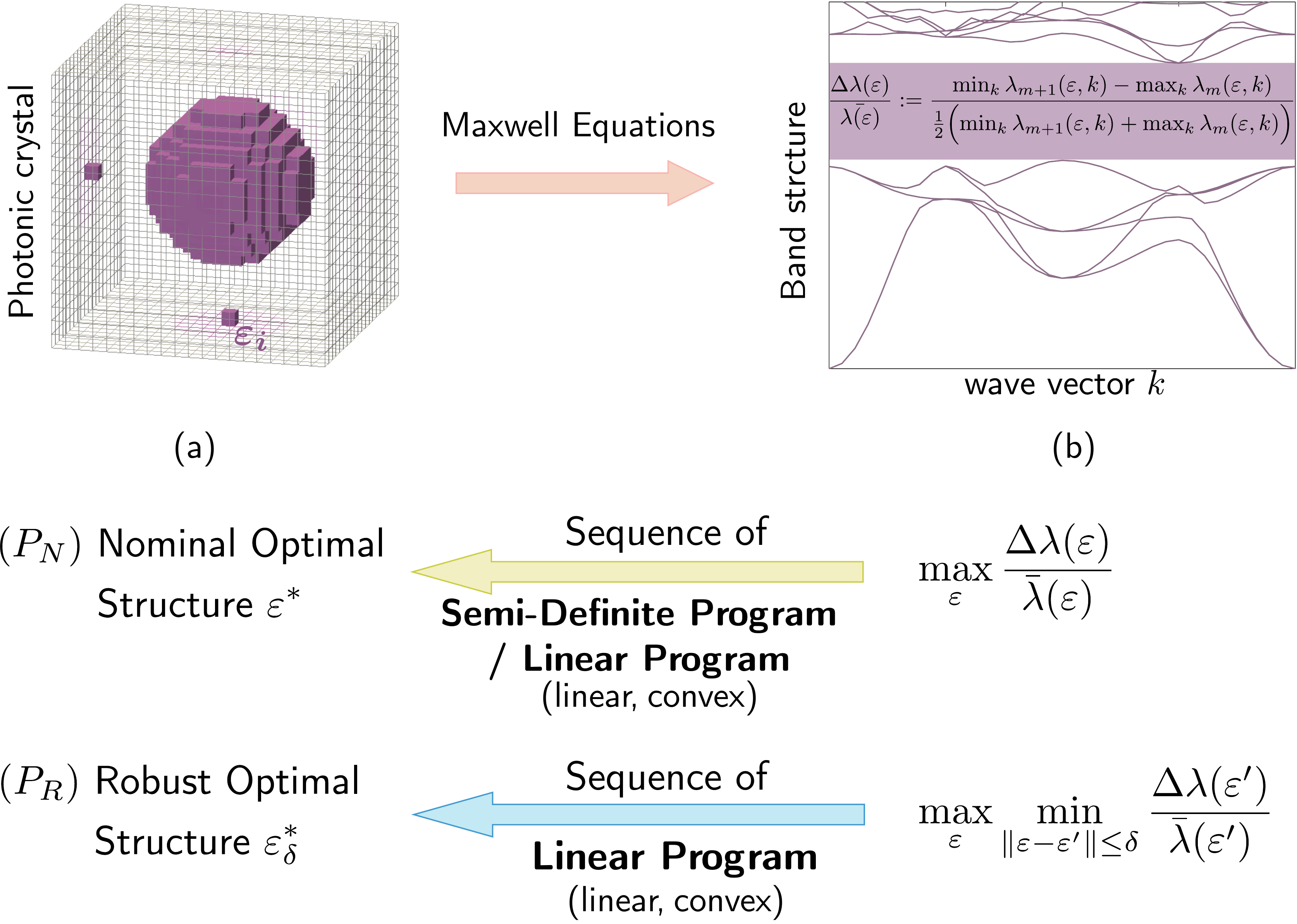}
\caption{Schematic of optimization process of photonic-crystal band-gap structures. Computation from (a) to (b) is often known as the \emph{forward problem}, in which given the photonic crystal dielectric function (or in a discrete representation $\varepsilon_i$), one computes the band structure by solving the Maxwell equations. The optimal design problem ($P_N$), or the \emph{inverse problem}, seeks to compute the optimal dielectric function $\varepsilon^*$ that maximizes the frequency band gap. The robust optimal design problem ($P_R$) seeks to compute a more robust optimal solution $\tilde{\varepsilon}_{\delta}^*$ by solving a \emph{maxmin} optimization problem, in case the nominal optimum $\varepsilon^*$ is not easily fabricable.}
\label{fig1}
\end{figure}

Photonic band gaps are frequency ranges $\Delta \omega$ in which there are no propagating electromagnetic waves, and have many potential applications, e.g, waveguides, filters, resonant cavities, etc. Band gaps can be achieved in a variety of periodic dielectric structures (photonic crystals) in 3D. The earliest 3D structures were designed by hand or by optimizing over a few parameters \cite{fan1994design, doosje2000pbo, biswas2002three, maldovan2002photonic, toader2003photonic, maldovan2003exploring, maldovan2004diamond, maldovan2005photonic, joannopoulos2011photonic}. In contrast to few-parameter optimization, topology optimization \cite{bendsoe2003topology} (in which the geometry is completely arbitrary, constrained only by the spatial resolution) involves qualitatively different computational methods and geometric parameterizations.  Previous topology parameterizations include level-set descriptions \cite{burger2004inverse, burger2005survey, kao2005mbg, he2007incorporating} and continuous relaxations in which $\epsilon$ (the dielectric permittivity) of every voxel can vary continuously in $[\epsilon_{\min}, \epsilon_{\max}]$ \cite{dobson1999maximizing, cox2000bso, shen2002large, sigmund2003systematic, jensen2004systematic, watanabe2006broadband, halkjaer2006maximizing, sigmund2008geometric, liang2013formulation}.  In principle, the latter approach can yield structures with unphysical intermediate materials, but we find in practice that this does not occur in our 3D band-gap optimization here (we encountered rare counter-examples elsewhere, e.g., $\S 4.3$ in \citeasnoun{men2010optimal}), and in any case there are various regularization approaches to eliminate these artifacts if needed (e.g., the homogenization method in \citeasnoun{bendsoe1988generating}, the SIMP method introduced in Refs. \citenum{bendsoe1995optimization} and \citenum{bendsoe1999material}, an alternative interpolation method proposed in \citeasnoun{stolpe2001alternative}, and the SINH method in \citeasnoun{bruns2005reevaluation}). A variety of optimization methods have been proposed for topology optimization of gaps \cite{dobson1999maximizing, cox2000bso, shen2002large, sigmund2003systematic, kao2005mbg} or fractional gaps \cite{halkjaer2006maximizing,sigmund2008geometric}. Although most previous works maximize the absolute gap $\Delta \omega$,  the fractional gap size $\Delta \omega/\bar{\omega}$ (where $\bar{\omega}$ is the mid-gap frequency) is typically the preferred metric \cite{joannopoulos2011photonic} due to its scale invariance, and optimizing $\Delta \omega$ generally produces a sub-optimal fractional gap. One major difficulty with frequency-gap optimization, and with eigenvalue optimization in general, is that eigenvalues are not generally differentiable functions of the design variables when eigenvalues are degenerate, and this can cause gradient-based optimization methods to break down. Moreover, we found that these breakdowns are particularly problematic in 3D band structures where many accidental degeneracies can arise. Methods like the generalized gradient have been proposed  to deal with the sensitivity of the repeated eigenvalues \cite{seyranian1994multiple,cox1995generalized}.  Our approach instead builds on a subspace projection to reformulate the optimization problem as a convex semidefinite program, in which eigenvalue sensitivities are not explicitly required. Schematically, the process is depicted in \figref{fig1}: we describe the unit cell by discretized $\varepsilon_i$ voxels (symmetrized in \secref{sec_methods}), solve it to find the band structure $\lambda(\vec{k},\varepsilon)$ ($\lambda = (\omega/c)^2$ are the Maxwell eigenvalues \cite{joannopoulos2011photonic}), and then optimize the fractional gap $f(\varepsilon) :=\Delta \lambda/ \bar{\lambda}$ by a sequence of SDPs. This nominal optimization problem is denoted by $P_N$. (We show in \secref{sec_methods} that optimizing $\Delta \lambda/ \bar{\lambda}$ is equivalent to optimizing $\Delta \omega/\bar{\omega}$, but we find the former more convenient. )

The optimization problem proposed above aims to find a \emph{nominal} optimal solution $\varepsilon^* = \arg \max_{\varepsilon} f(\varepsilon)$ (where $\varepsilon\in \mathbb{R}^n$ parameterizes the geometry), in which the process assumes the fractional gap $f$ is exact and deterministic, and the optimal solution $\varepsilon^*$ can be fabricated precisely. This clearly poses limitations in realistic settings where $f$ may contain uncertainties or $\varepsilon^*$ cannot be exactly built due to fabrication errors. A methodology to address this problem is \emph{robust} optimization, which is a broad category of optimization approaches that take uncertainties into account \cite{beyer2007robust, ben2009robust, bertsimas2010robust, bertsimas2011theory}. Here, we adopt a \emph{maximin} version of the robust optimization formulation, see $P_R$ in \figref{fig1}, in which the worst case is optimized, i.e., $\tilde{\varepsilon}^* = \arg \max_{\varepsilon} \min_{\varepsilon'} f(\varepsilon, \varepsilon')$, and the variable $\varepsilon'$ contains the uncertainties. Previous works on robust optimization either focused on convex robustification where the robust formulation retains the convex structure of the nominal problem \cite{ben2009robust,bertsimas2010robust, bertsimas2011theory}, or on nonconvex robust-optimization problems with the worst-case unknowns $\varepsilon'$ residing in a space independent of $\varepsilon$ \cite{mutapcic2009robust, sigmund2009manufacturing, wang2011robust, wang2011projection, schevenels2011robust, oskooiMu2012robust, elesin2012design}. In the framework of topology optimization, in which each voxel is an independent degree of freedom, one important type of uncertainty is in the value of each voxel, i.e., $\varepsilon, \epsilon'$ are in the same space $S$. The robust formulation hence becomes $\tilde{\varepsilon}_{\delta}^* = \arg \max_{\varepsilon\in S} \min_{\varepsilon'\in S, \|\varepsilon'-\varepsilon\|\le \delta} f(\varepsilon, \varepsilon')$. The modeling and computational issues of this formulation were addressed in 2D photonic crystals by our previous work \cite{men2014fabrication}. 

\section{Problem Formulations and Solution Methods} 
\label{sec_methods}

The time-harmonic Bloch-wave magnetic-fields $\vec{H}(\vec{r}) e^{i\vec{k}\cdot \vec{r} - \omega t}$ of the Maxwell equations solve the following eigenproblem\cite{joannopoulos2011photonic},
\begin{equation}
\nabla_{\vec{k}} \times \left(\frac{1}{\varepsilon(u)} \nabla_{\vec{k}} \times \vec{H} \right) = \Bigl( \frac{\omega}{c} \Bigr)^2 \vec{H} := \lambda \vec{H},
\label{eigeq}
\end{equation}
where $c$ is the vacuum speed of light, and $\nabla_{\vec{k}}  := \nabla + i \vec{k}$ for each Bloch wavevector $\vec{k}$ in Brillouin zone $\mathcal{B}$. In the nominal design problem $P_N$, we seek to maximize the \emph{ frequency-gap ratio} $g(u)$ between two bands $\omega_m(\varepsilon(u), \vec{k})$ and $\omega_{m+1}(\varepsilon(u), \vec{k})$,
\begin{equation}
\displaystyle g(u) := \frac{\underset{\vec{k} \in \mathcal{B}}{\min} \omega_{m+1} (\varepsilon(u), \vec{k}) - \underset{ \vec{k} \in \mathcal{B}}{\max}\omega_{m} (\varepsilon(u), \vec{k}) }{\frac{1}{2} \left(\underset{\vec{k} \in \mathcal{B}}{\min} \omega_{m+1} (\varepsilon(u), \vec{k}) + \underset{\vec{k} \in \mathcal{B}}{\max}\omega_{m} (\varepsilon(u), \vec{k}) \right) }.
\label{freqgap}
\end{equation}
The simplest parametrization of $\varepsilon$, given a computational solver for \eqref{eigeq} on a discrete grid, would simply be $\varepsilon_i \in [\varepsilon_{\min}, \varepsilon_{\max}]$ on each grid point. Our implementation is similar to this approach in spirit, but uses a slightly different parametrization $\varepsilon(u)$ for variables  $u\in \mathcal{D} := [0,1]^{N_u}$ that is described in detail below. 

It turns out that the fractional gap in eigenvalues $\lambda = (\omega/c)^2$,
\begin{equation}
 f(u) := \frac{\underset{\vec{k} \in \mathcal{B}}{\min} \lambda_{m+1} (\varepsilon(u), \vec{k}) - \underset{\vec{k} \in \mathcal{B}}{\max}\lambda_{m} (\varepsilon(u), \vec{k}) }{\frac{1}{2} \left(\underset{ \vec{k} \in \mathcal{B}}{\min} \lambda_{m+1} (\varepsilon(u),  \vec{k}) + \underset{\vec{k} \in \mathcal{B}}{\max}\lambda_{m} (\varepsilon(u),  \vec{k}) \right) }
\label{opt}
\end{equation}
is a monotonic function of $g(u)$; hence it is equivalent to optimize $f(u)$ or $g(u)$, and we found $f(u)$ easier to work with. It is a straightforward exercise to prove this monotonicity, e.g., by showing $[ g(u_1)-g(u_2) ][ f(u_1)-f(u_2) ] \ge 0$ for all $u_1, u_2\in \mathcal{D}$ from the semi-definiteness \cite{joannopoulos2011photonic} of the eigenproblem. Conversely, neither $g$ nor $f$ are monotonic in the absolute gap $\Delta\omega$, so our optimization problem is \emph{not} equivalent to optimizing $\Delta \omega$.

In principle, the band gaps must be determined from the band extrema over the entire Brillouin zone $\mathcal{B}$.    However, the problem simplifies for the high-symmetry structures to which we constrain ourselves below. First, $\mathcal{B}$ can be reduced by the rotational/mirror symmetries to an irreducible Brillouin zone (IBZ) \cite{joannopoulos2011photonic}.   Furthermore, it can easily be shown that the vertices and edges of the IBZ (denoted as $\partial \mathcal{B}$) are local extrema of the bands, due to the rotational symmetries.  Empirically, it has been observed that the \emph{global} band extrema almost always fall on $\partial \mathcal{B}$, with rare exceptions \cite{harrison2007occurrence}.  As a practical matter, one typically designs photonic band gaps by considering only $\partial \mathcal{B}$, and then the interior of $ \mathcal{B}$ is checked \emph{a posteriori} \cite{joannopoulos2011photonic}. We adopt that procedure here, as well: we only consider $\partial \mathcal{B}$ when maximizing $f(u)$, and we verify afterwards that none of the optimized structures have gap edges elsewhere in $\mathcal{B}$.

The optimal solution of $\max_u f(u)$ often cannot be fabricated directly, for example, due to fabrication errors, technological limitations on the resolution of the fine features, or unavailable materials as a result of a non-binary solution $u$, i.e., $0 < u_{i\in \mathcal{I}} < 1$. This nonfabricability of the nominal optimal design occurs in many optimization algorithms that do not explicitly incorporate in the solution robustness, and is an especially common issue in topology optimization, because the large number of degrees of freedom may make it easier to find non-robust solutions. However, large band gaps tend to be inherently robust to non-systematic errors, e.g., surface roughness \cite{rodriguez2005disorder}, and so we found that it was not necessary to design that type of robustness explicitly. Instead, we modified our algorithm to ensure robustness to \emph{systematic} (i.e., periodic and symmetric) errors, by solving the following \emph{maximin} fabrication-adaptivity (FA) problem~\cite{men2014fabrication}:
\begin{equation}
\tilde{u}^*_{\delta} = \arg \max_{u\in[0,1]^{N_u}} \ \ \min_{\| u' - u\|_1/N_u \le \delta, \ u'\in[0,1]^{N_u}}  \quad f(u'),
 \label{opt_fa}
\end{equation}
where $\delta$ is the robustness parameter and $\| \cdot \|_1$ denotes the $L_1$ norm, and signifies the percentage range of allowable perturbation.

The basic optimization algorithm used here for solving \eqref{opt} is a subspace-based semidefinite-program (SDP) formulation previously proposed in \citeasnoun{men2010bandgap}. The FA optimization problem in \eqref{opt_fa} is solved using a linear-program (LP) based iterative algorithm proposed in $\S$3 of \citeasnoun{men2014fabrication}. Numerical solutions to the Maxwell eigenvalue equation \eqref{eigeq} required by both optimization problems $P_N$ and $P_R$ are computed using an efficient preconditioned block-iterative eigensolver in a planewave basis (from the MPB package \cite{johnson2001block}). 

\begin{figure}[h]
\centering
\includegraphics[scale = 0.3]{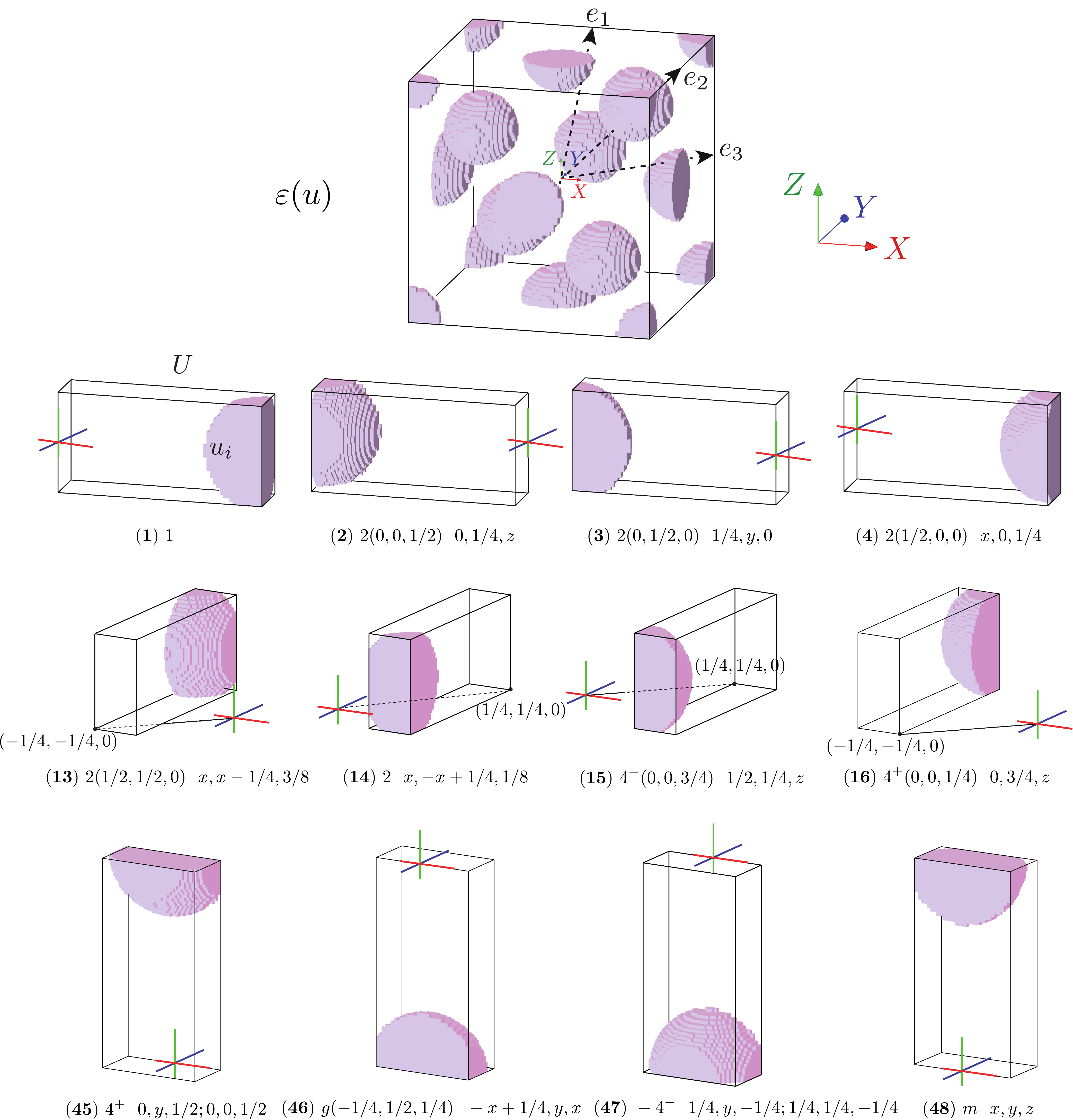}
\caption{Material structure and the optimization design region in a diamond lattice. The optimization design region is limited to the rectangular parallelepiped, denoted by $U$ in $(1)$, and is only $1/64$ of the cube. Through necessary symmetric operations, e.g., $(2), \ldots, (48)$, the material $\varepsilon(u)$ (top) is reconstructed.}
\label{fig_sg}
\end{figure}

Successful optimization of the 3D structures also relies on some efficient geometry modeling. The periodic unit cell in 3D photonic crystal is a parallelepiped, but in order to restrict ourselves to high-symmetry structures  (e.g. the diamond symmetry in \figref{fig_sg}), we should, in principle, have design variables $\varepsilon_i$ only in the ``asymmetric unit''  \cite{hahn2006international} of the cell: this is a typically wedge-shaped polyhedron whose symmetry operations (e.g., rotations, reflections, etc.) yield the entire unit cell.   However, it is computationally inconvenient to define a structured mesh of design variables over such a polyhedron. Instead, we employ the following transformation.  We define a 3D grid of design variables $u_i \in [0,1]$ over the smallest parallelepiped U that encloses the asymmetric unit, as shown at the upper left of Fig. 2.  To obtain the material $\varepsilon$ at any point $\vec{r}$, we first perform all symmetry operations on U, interpolate $u(\vec{r})$ from the grid $u_i$ for each of these transformations (e.g., 48 for diamond symmetry), compute the \emph{average} $\bar{u}$ of all of these $u(\vec{r})$, and finally obtain $\epsilon(\vec{r}) = \epsilon_{\min} + \bar{u} (\epsilon_{\max} - \epsilon_{\min})$.   This projection procedure allows us to easily impose any desired symmetry group while maintaining the simplicity of a Cartesian grid of unknowns.

Besides the geometry modeling discussed above, we also explore efficient parallel computation whenever possible. The eigensolver MPB \cite{johnson2001block} includes a distributed-memory parallel version. The optimization solver we use for the SDP and LP problems, MOSEK\cite{mosek2013mosek}, supports multithreaded computation. In addition, the FA problem \eqref{opt_fa} is also conveniently parallelized, thanks to the numerous and independent LPs to be solved. The resulting software allow us to solve a typical 3D problem $P_N$ in \secref{subsec_opt} in less than an hour with $8$ CPUs, and a typical 3D problem $P_R$ in \secref{subsec_fa} in less than a day with $8$ CPUs.

\section{Results and Discussion}
\label{sec_results}
For most of the structures shown below (except for \secref{subsec_idx}), we consider two contrasting dielectric materials, where the high refractive-index material has $n_{hi} = 3.6$ (similar to GaAs in the near infrared), and the low refractive-index material has $n_{lo}= 1$ (air). We also consider various prescribed symmetries, for example, simple cubic (space group no. 221), face-centered cubic (no. 225), body-centered cubic (no. 214) and diamond (no. 227), as they correspond to some known structures with sizable band gaps \cite{biswas2002three,maldovan2004diamond,maldovan2005photonic}.

\subsection{Optimal structures: SC5, Diamond2, Diamond8, BCC2, and FCC8} 
\label{subsec_opt}
Starting from any random distribution of material configuration (which generically has a negative gap $f$), the optimization algorithm is always able to increase the gap size. However, we are only able to obtain gaps between certain pairs of bands ($m$ and $m+1$), similar to previous authors (and unlike 2D \cite{kao2005mbg,sigmund2008geometric,men2010optimal}). Furthermore, the non-convexity means that some starting points lead to suboptimal local optima (often small or negative gaps) even for the known separable bands. Nevertheless, by repeating the optimization for many random starting points, we were able to obtain several structures with large gaps. For the diamond structures, we only found a handful of local optima, so that the large-gap structure was found for about $20\%$ of starting points. For the simple-cubic structures, the large-gap structure was found for only about $5\%$ of the starting points. For the face-centered cubic structures, we find at least two optima, one inverse-opal--like structure about 5\% of the time and another (larger-gap) structure about 15\% of the time.   As explained in \secref{sec_methods}, we optimize one symmetry group at a time, where constraining the symmetry group has the essential benefit of allowing us to evaluate only the edges of the irreducible Brillouin zone.

\begin{figure}
\centering
\subfigure[SC5 (no. 221)]{
\includegraphics[scale=0.2]{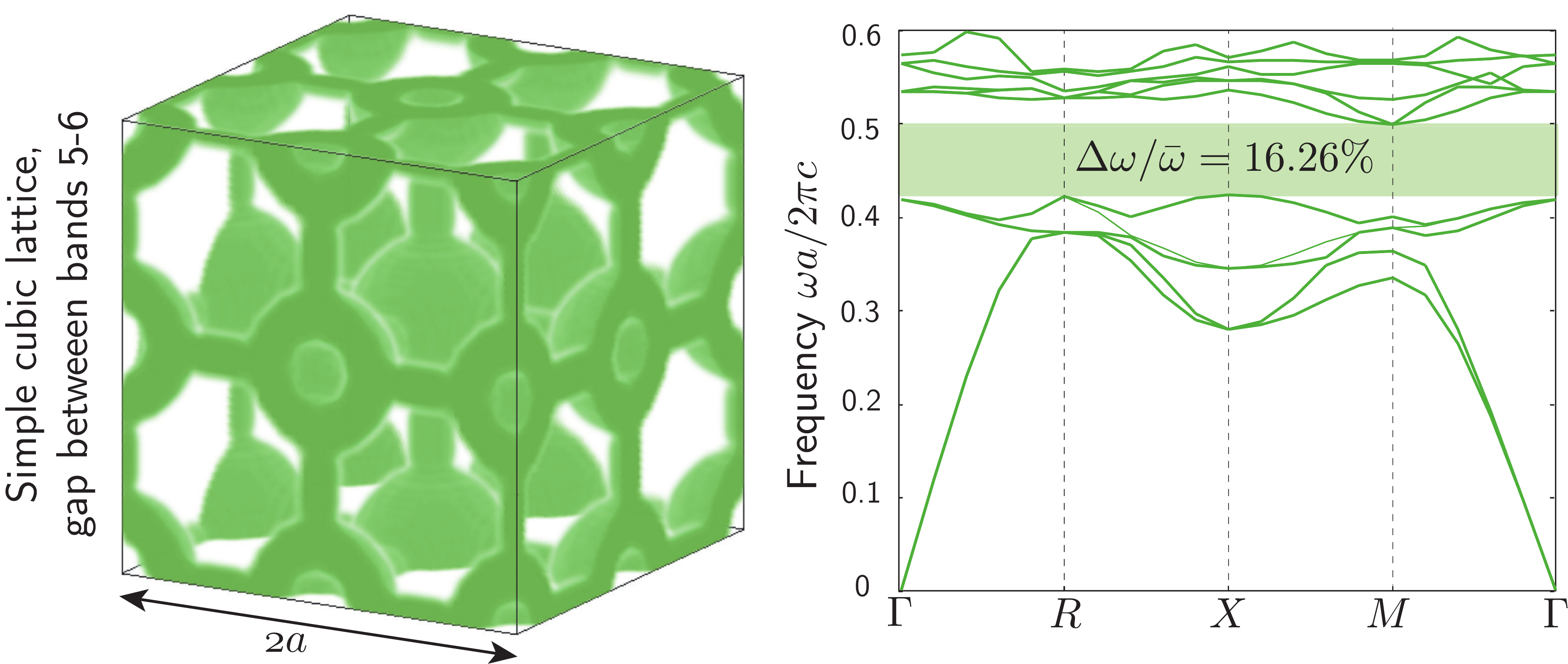}}
\subfigure[Diamond2 (no. 227)]{
\includegraphics[scale=0.2]{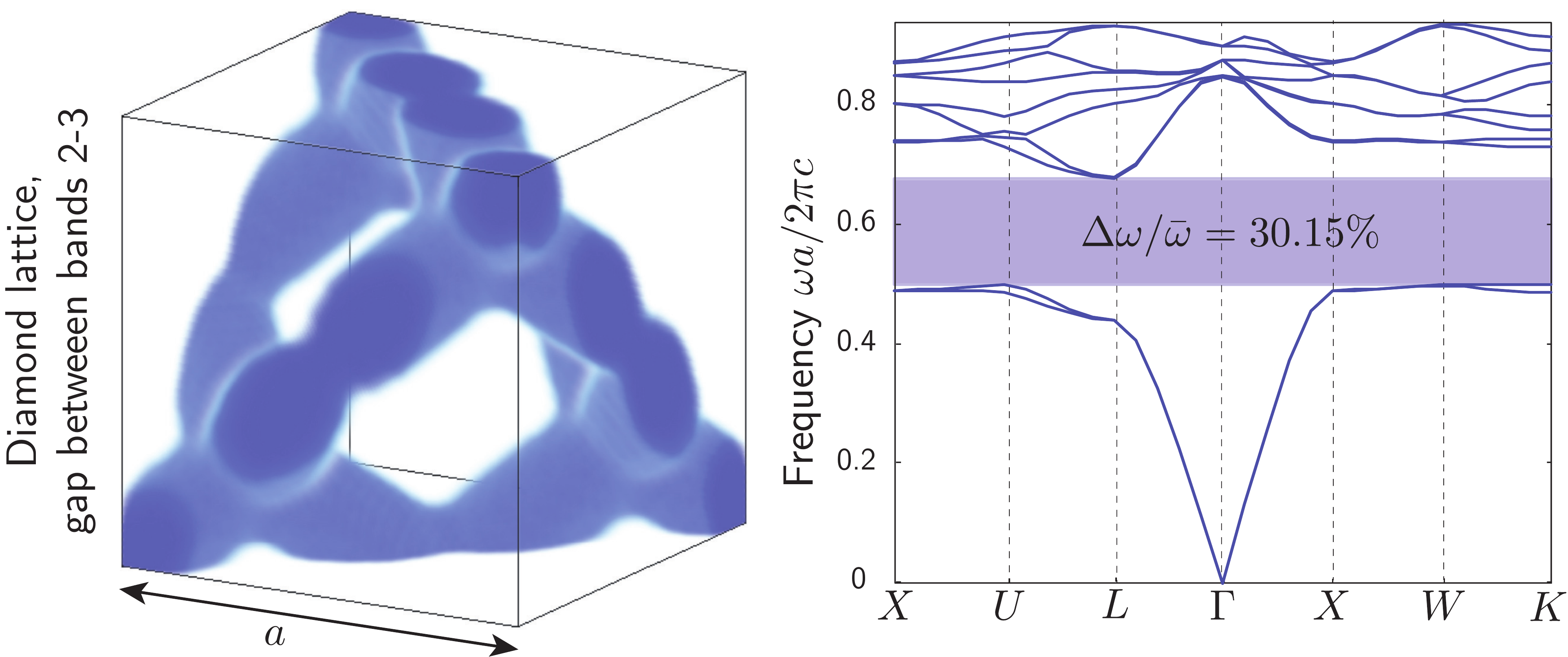}}
\subfigure[Diamond8 (no. 227)]{
\includegraphics[scale=0.2]{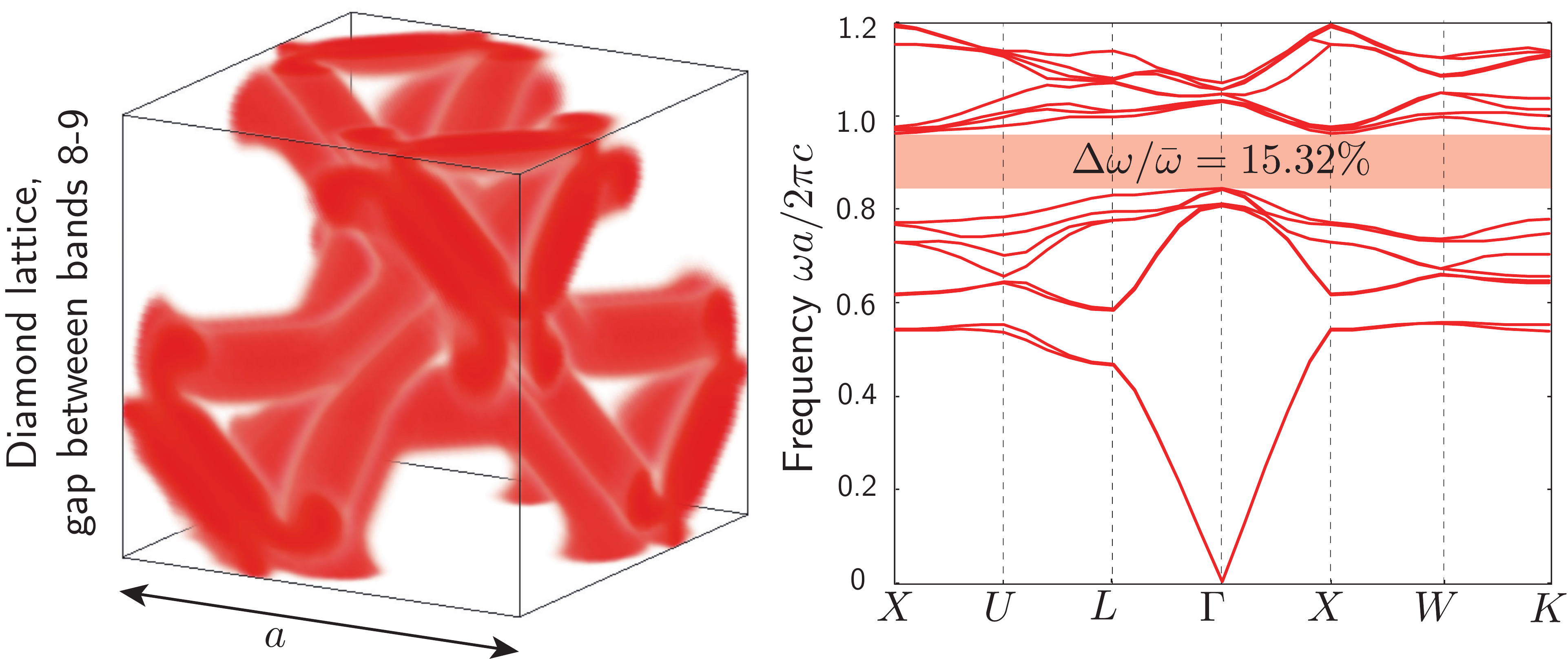}}
\subfigure[BCC2 (no. 214)]{
\includegraphics[scale=0.2]{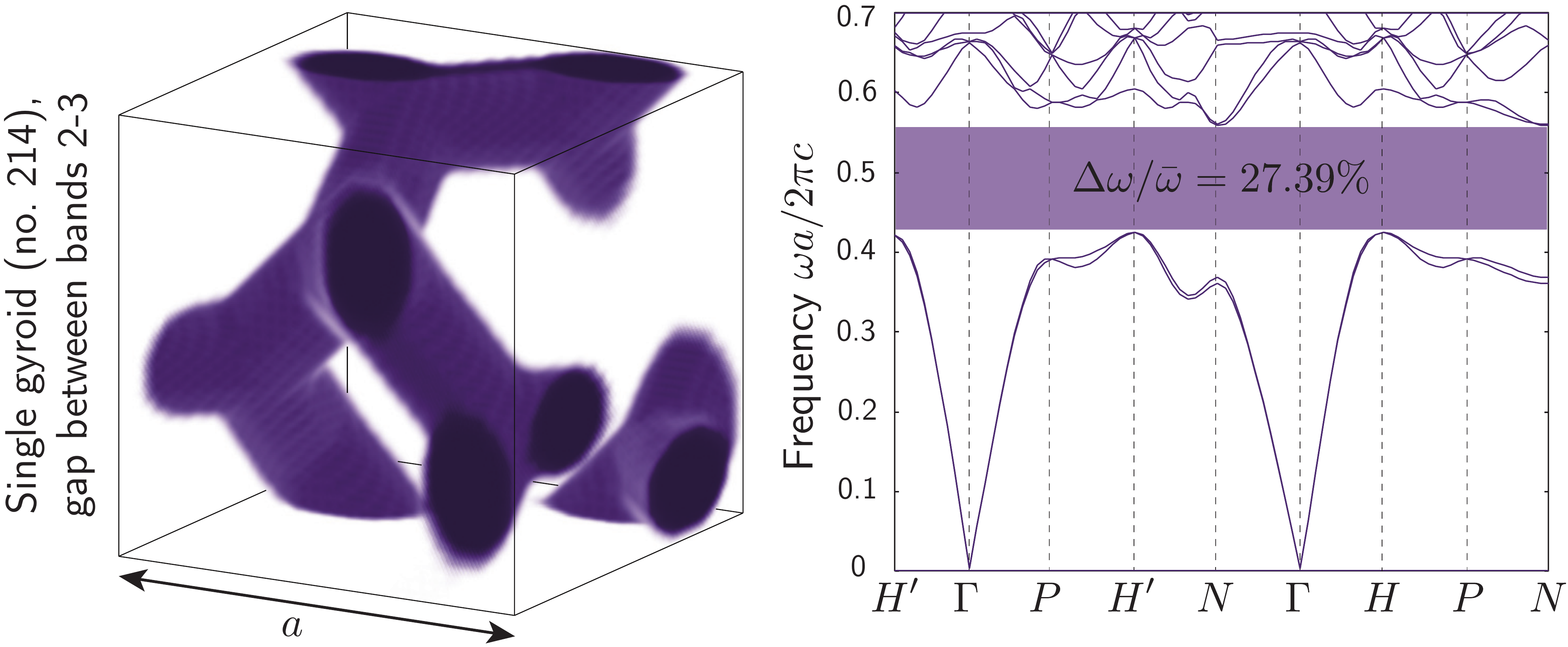}}
\subfigure[FCC8 (no. 225)]{
\includegraphics[scale=0.2]{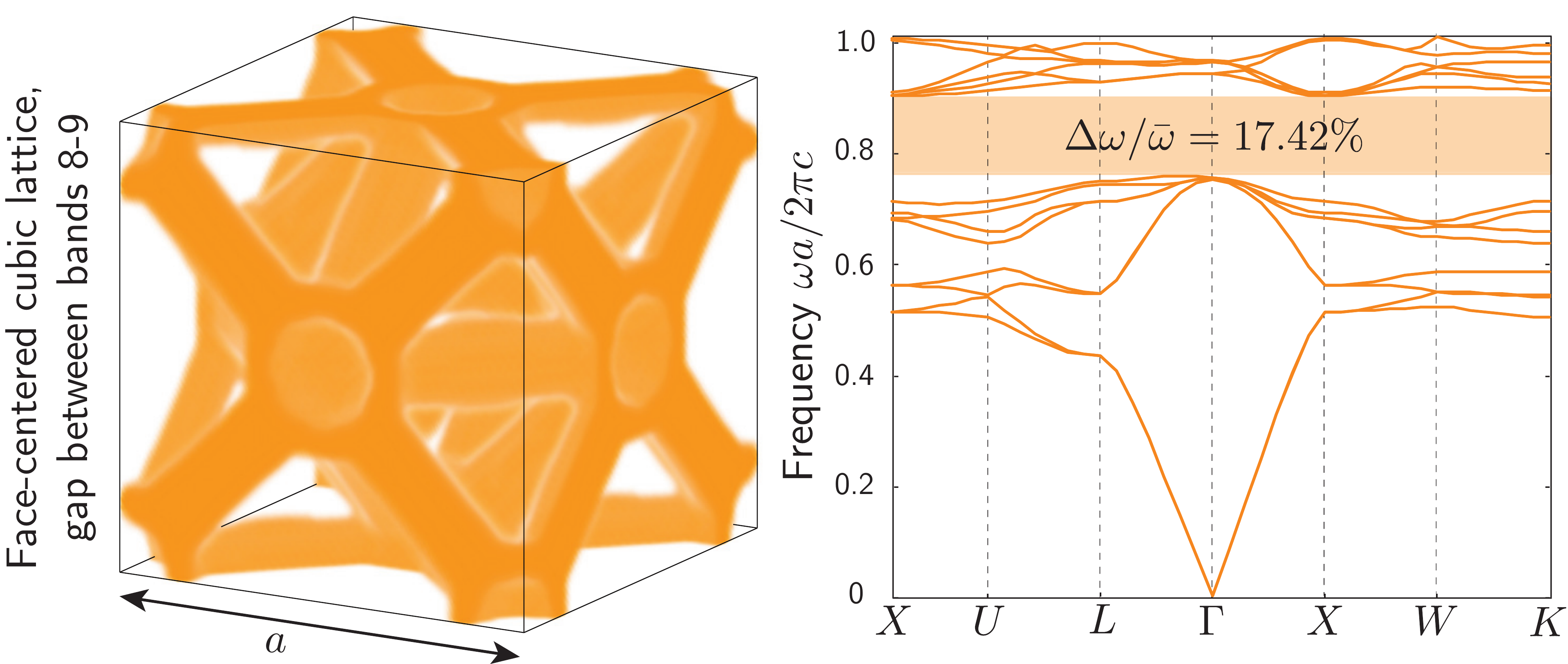}}
\caption{Optimized photonic crystals of prescribed symmetries (space groups no. 214, 221, 225, 227) with complete gaps between consecutive bands.}
\label{fig_optima}
\end{figure}

A structure constrained to have symmetry $Pm\bar{3}m$ (space group no. 221) in simple cubic lattice, denoted ``SC5'', is shown in \figref{fig_optima}(a) and resembles a cubic lattice of hollow spheres connected by cylindrical ``bonds.'' SC5 has a frequency gap of $16.26\%$ between the $5$th and the $6$th bands.  Structures very similar to SC5, which also had a gap of $\sim 13\%$ (for $3.6:1$ index contrast) between the same bands, were found by few-parameter optimization in previous work~\cite{biswas2002three,maldovan2002photonic,maldovan2005photonic}, although the previous work did not identify the possibility of improving the gap by introducing air holes in the center of each dielectric sphere. SC5 was the only case of a substantial gap ($>10\%$) in a cubic lattice that we found after examining many pairs of bands, and is probably quite challenging to fabricate at infrared scales.

A structure constrained to have symmetry $Fd\bar{3}m$ (space group no. 227) in face-centered cubic lattice (or commonly known as diamond symmetry), denoted ``Diamond2'', is shown in \figref{fig_optima}(b), and has a $30.15\%$ gap between the $2$nd and $3$rd bands.  The topology of this design is very reminiscent of a large number of diamond-like photonic-crystal hand designs in previous works~\cite{maldovan2004diamond}, all of which had 20--30\% gaps between the second and third bands.   (The connectivity resembles the bond pattern in atomic diamond structures~\cite{joannopoulos2011photonic}.)   The many published variations on this structure were intended to adapt the structure to various fabrication technologies.  Our results show that, even if one removes the constraint of easy fabrication, only a slight improvement on the gap size can be obtained for this symmetry and this pair of bands.

We also examined many other pairs of bands in diamond-symmetry structures, and only found one other possibility for a large gap, which is denoted ``Diamond8'' and is shown in \figref{fig_optima}(c).  This structure has a  $15.32\%$ complete gap between the 8th and 9th bands, and does not bear any obvious resemblance to previous designs. Since the gap is smaller than in Diamond2, however, there seems little reason to seek a variant of Diamond8 that might be easier to fabricate.

A structure constrained to have symmetry $I4_132$ (space group no. 214) in body-centered lattice, denoted ``BCC2'', is shown in \figref{fig_optima}(d), and has a gap of $27.39\%$ between the $ 2 $nd and $ 3 $rd bands. The existence of this gap has been previously reported in \cite{chutinan1998spiral,maldovan2002photonic}, but with smaller sizes than the optimum we have here. The crystal structures correspond to a family of bicontinuous cubic structures, specifically the \emph{single gyroid}, which have been explored for self-assembly of large-scale photonic materials. 

\begin{figure}
\centering
\subfigure[Initial solution and local optimum No. $1$]{
\includegraphics[scale=0.25]{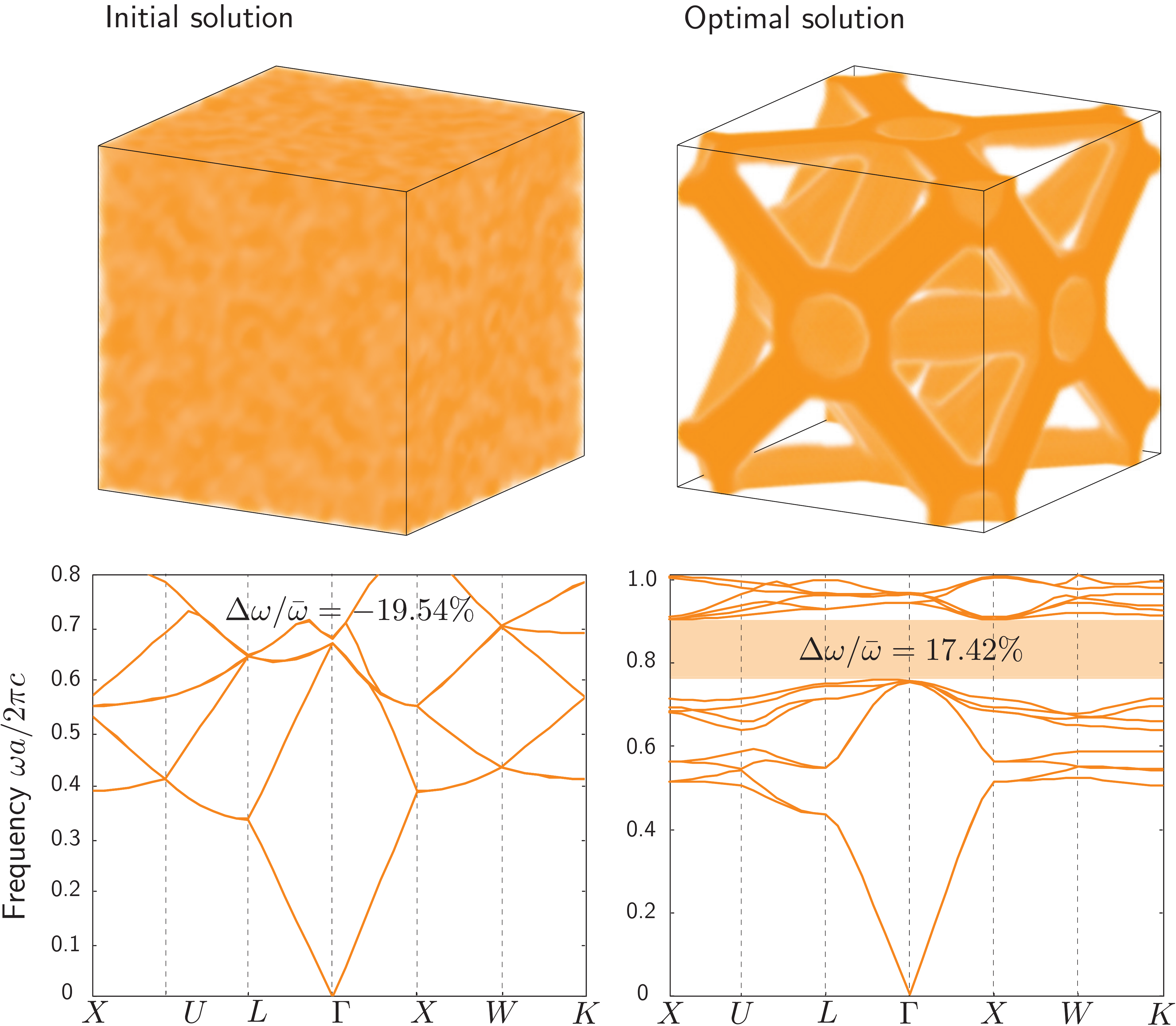}}
\\
\vspace{1cm}
\subfigure[Initial solution and local optimum No. $2$]{
\includegraphics[scale=0.25]{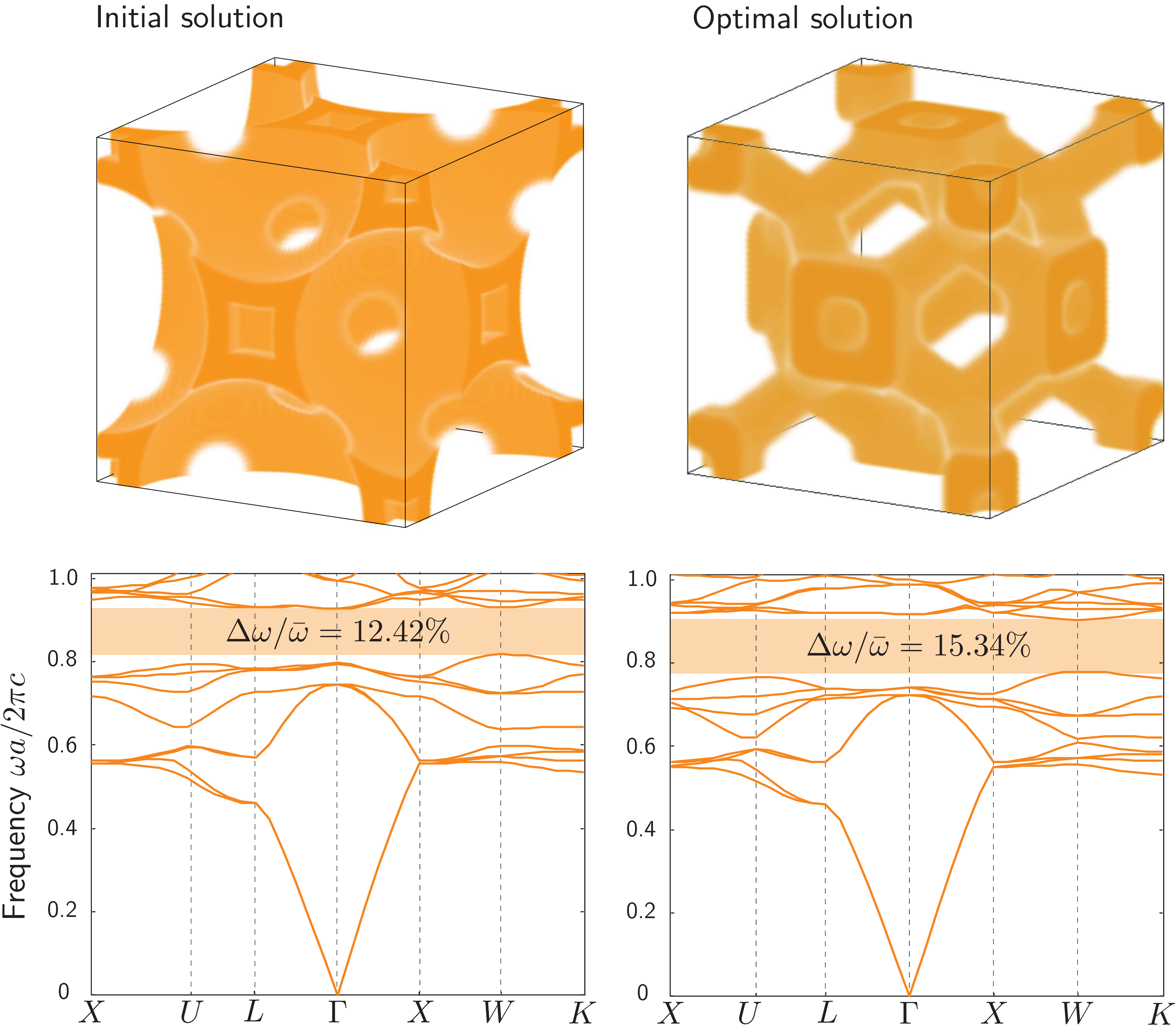}}
\caption{Band structures of initial and optimized photonic crystals of a face-centered cubic (fcc) lattice (no. 225). The two initial solutions and their band structures are shown on the left, and their corresponding optimal solutions and the band structures are shown on the right.}
\label{fig_fcc8ba}
\end{figure}

A structure constrained to have symmetry $Fm\bar{3}m$ (space group no. 225) in face-centered cubic lattice, denoted ``FCC8'', is shown in \figref{fig_optima}(e), and has a gap of $17.42\%$ between the $8$th and $9$th bands. As we go to higher bands and lower-symmetry structures, such as FCC8, we find that there are many more local optima in the gap-optimization problem. For example  in \figref{fig_fcc8ba}, two locally optimal structures are obtained starting from different initial solutions. With a random initial structure (initial $u$) like the one shown on the top left of \figref{fig_fcc8ba}(a), which has a negative gap, we often obtain the optimal crystal structure shown on the top right. About 5\% of the time, however, or alternatively if we had started the optimization with the  ``inverse opal'' structure (top left of \figref{fig_fcc8ba}(b))~\cite{sozuer1992photonic} (a close-packed fcc lattice of air holes in a dielectric matrix), we would obtain an inverse-opal--like local optimum shown at the top right of \figref{fig_fcc8ba}(b) with a gap of $15.35\%$.   (Note that optimization also reproduces the later discovery that it is advantageous to introduce additional air voids and pores \cite{busch1998photonic} which can be modeled by overlapping spherical shells \cite{joannopoulos2011photonic}.)   The FCC8 structure seems to be topologically distinct from the inverse opal: if we view the hollow dielectric blobs at the faces and corners of the cubes as ``atoms,'' then the atoms in the inverse opal are connected by 8 rod-like ``bonds'' per atom, whereas the FCC8 structure has 12 bonds per atom.  Essentially, FCC8 is an fcc lattice of small hollow spherical shells, each of which is connected to all 12 nearest neighbors via dielectric rods (and we explicitly reparameterize the structure in this fashion below).

\subsection{Gap vs. geometry parametrization: SC5$(r_1, r_2, r_3)$, Diamond2$(r)$ and FCC8$(r_1, r_2, r_3)$}

\label{subsec_geo}
\begin{figure}
\centering
\includegraphics[scale = 0.3]{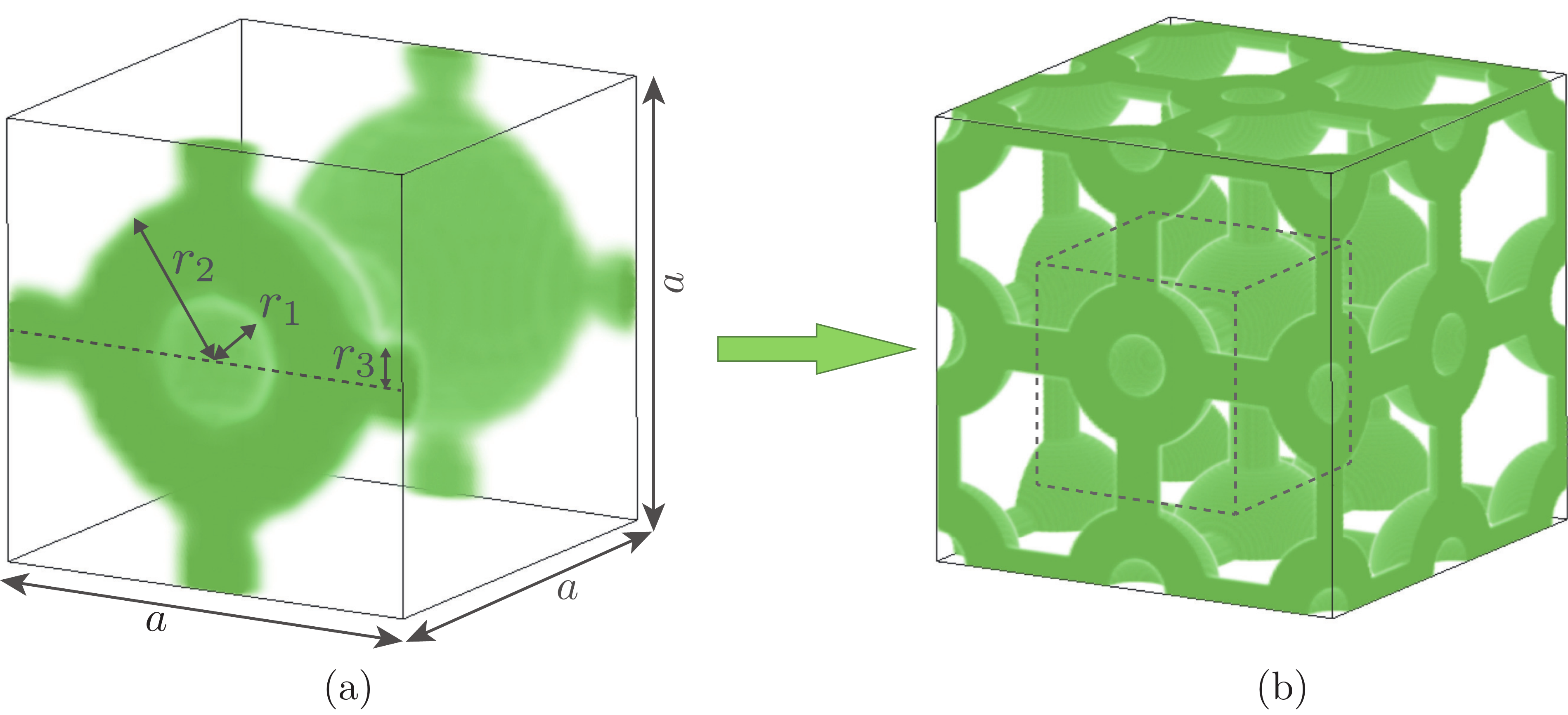}
\caption{(a) Geometry characterization of SC5 with three parameters $(r_1, r_2, r_3)$. (b) Reconstructed photonic crystal with $(r_1^{\ast}, r_2^{\ast}, r_3^{\ast})=(0.14a, 0.36a, 0.105a)$, and a frequency gap of $17.48\%$.}
\label{fig_sc5_geometry}
\end{figure}

\begin{figure}
\centering
\includegraphics[scale = 0.3]{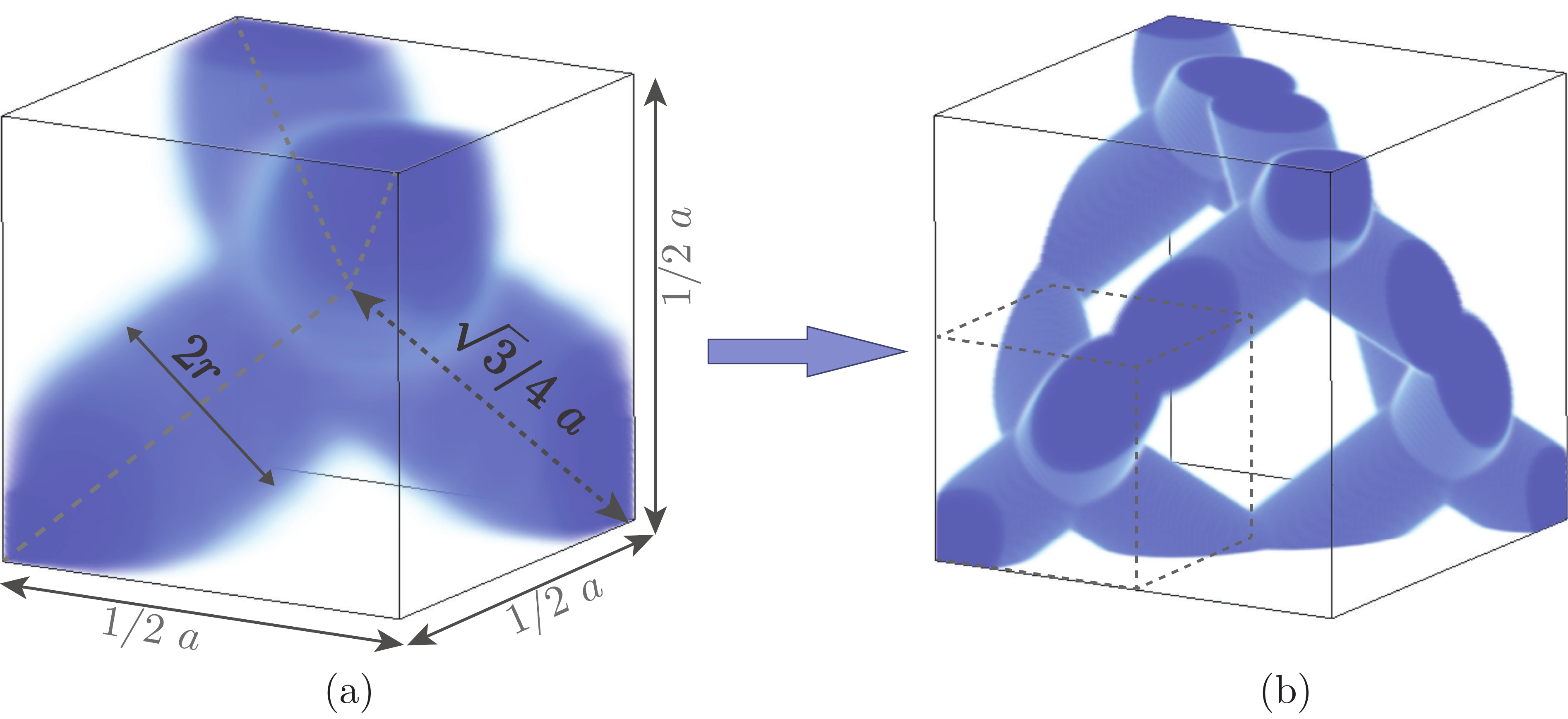}
\caption{(a) Geometry characterization of Diamond2 with one parameters $r$. (b) Reconstructued photonic crystal with $r^{\ast} = 0.1a$, and a frequency gap of $31.56\%$.}
\label{fig_d2_geometry}
\end{figure}

\begin{figure}
\centering
\includegraphics[scale = 0.3]{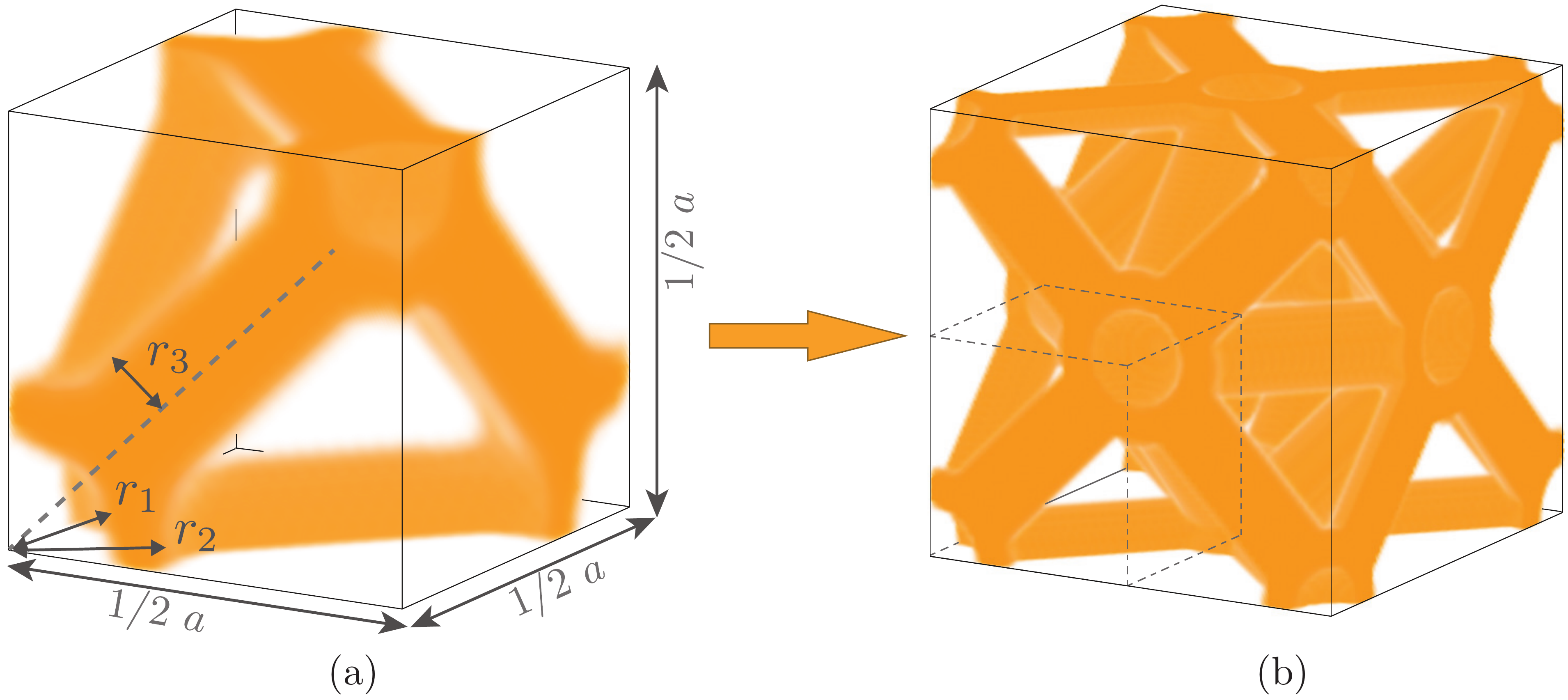}
\caption{(a) Geometry characterization of FCC8 with three parameters $(r_1, r_2, r_3)$. (b) Reconstructued photonic crystal with $(r_1^{\ast}, r_2^{\ast}, r_3^{\ast})=(0.12a, 0.19a, 0.08a)$, and a frequency gap of $18.30\%$.}
\label{fig_fcc8_geometry}
\end{figure}

Although we used topology optimization to exploit the maximum number of degrees of geometric freedom, the resulting SC5 and Diamond2 structures are relatively simple and can ultimately be described by a much smaller number of parameters, and the gaps are relatively insensitive to small geometric variations (such as the exact shape of the diamond ``bonds''). For example, we can reproduce a structure very similar to SC5 with a sphere-and-cylinder structure characterized by three parameters $(r_1, r_2, r_3)$, shown in \figref{fig_sc5_geometry}(a). Here $r_1$ denotes the radius of the air sphere, $r_2$ denotes the radius of the dielectric sphere, and $r_3$ denotes the radius of the dielectric cylinder of length $a-2r_2$, where $a$ is the lattice constant. Optimizing over those three parameters, with values given in \figref{fig_sc5_geometry}(a), yields a gap that is actually slightly larger ($\sim 17\%$) than the one discovered by topology optimization ($\sim 15\%$), but this appears to be an artifact of the finite spatial resolution---even at the same spatial resolution, the MPB software uses a more accurate subpixel-averaging technique~\cite{johnson2001block} in the case of simple geometric shapes (sphere and cylinders) than for the grid-based parameterization of topology optimization.   Similarly, Diamond2 can be roughly reproduced by a diamond lattice of dielectric cylinder ``bonds`` parameterized by their radius $r$, shown in \figref{fig_d2_geometry}(a). For a radius $r^{\ast} = 0.1a$, this reconstructed structure has a $31.56\%$ gap, versus the $30.15\%$ of the optimal solution, and again the difference seems to be an artifact of the finite spatial resolution and different subpixel averaging techniques in the two simulations.   FCC8 can also be reconstructed by a similar sphere-and-cylinder structure parameterized by three parameters $(r_1, r_2, r_3)$, shown in \figref{fig_fcc8_geometry}(a). The main advantage of this kind of re-parameterization of the topology-optimized structure is that it makes the results easier to communicate: anyone can use the reconstructed parameters to reproduce our results, whereas reproducing the topology-optimized structure would be difficult without access to the electronic data files.

\subsection{Gap vs. refractive index contrast $n_{hi}/n_{lo}$: SC5, Diamond2, and FCC8}
\label{subsec_idx}
\begin{figure}
\centering
\includegraphics[scale = 0.65]{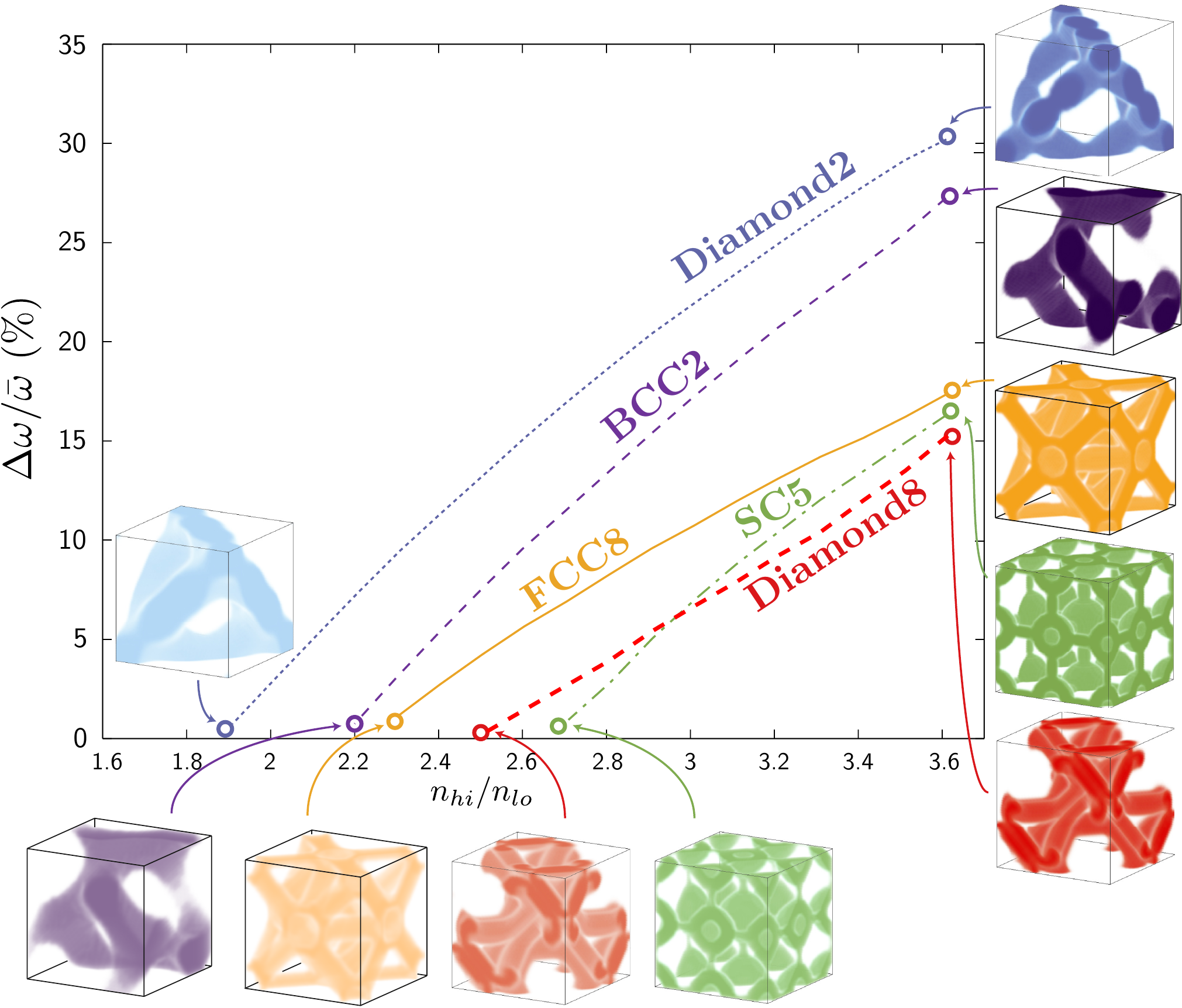}
\caption{Gap vs. refractive-index contrast $n_{hi}/n_{lo}$}
\label{fig_gap_idx}
\end{figure}

Photonic band-gap sizes are known to be dependent on the refractive-index contrast of the two constituent dielectric materials. For similar structures, one generally expects the optimal gap to reduce monotonically as the index contrast decreases (e.g. see Appendix C in \citeasnoun{joannopoulos2011photonic}), although the optimal parameters will also change with index contrast.  An important question is the minimum index contrast for which a band gap is possible.  To answer this and similar questions, we plot the optimal gap as a function of index contrast in \figref{fig_gap_idx} for different symmetries and pairs of bands.  The smallest index contrast for which we found a gap was $n_{hi}/n_{lo} = 1.9$, for the Diamond2 structure, very similar to previous results for hand design of this type of structure\cite{joannopoulos2011photonic,johnson2000three,maldovan2004diamond,ho1990existence}. The optimal structures at the highest and lowest index contrasts are shown as insets of \figref{fig_gap_idx}, and the general trend is that at higher index contrasts, the dielectric veins become thinner (a well known phenomenon dating back to quarter-wave stacks in 1D \cite{joannopoulos2011photonic}) and more regular (cylindrical or spherical) in shape.

\subsection{Gap vs. FA parameter: $Diamond2(\delta)$}
\label{subsec_fa}

As discussed in Sec.~\ref{sec_methods}, we also considered a robust/fabrication-adaptive (FA) version of our gap-optimization problem, which maximizes the worst-case gap with respect to uncertainty in the parameters $u$.   In this section, we present results from FA optimization of the Diamond2 structure as a function of the amount of uncertainty $\delta$ (the mean absolute error in $u(\vec{r})$ at each point).  The $\delta = 0$ solution is equivalent to the original (``nominal'') optimization problem, and with increasing $\delta$ one expect the optimum worst-case to have a smaller gap.   However, if one simply added uncertainty to the nominal design, one would expect its gap to decrease faster with $\delta$ than the FA designs (which are redesigned for each $\delta$).   As discussed in Sec.~\ref{sec_methods}, increasing $\delta$ represents increasing \emph{systematic} errors in the structure (i.e. errors that are periodic and symmetrical).

\begin{figure}
\centering
\includegraphics[scale = 0.6]{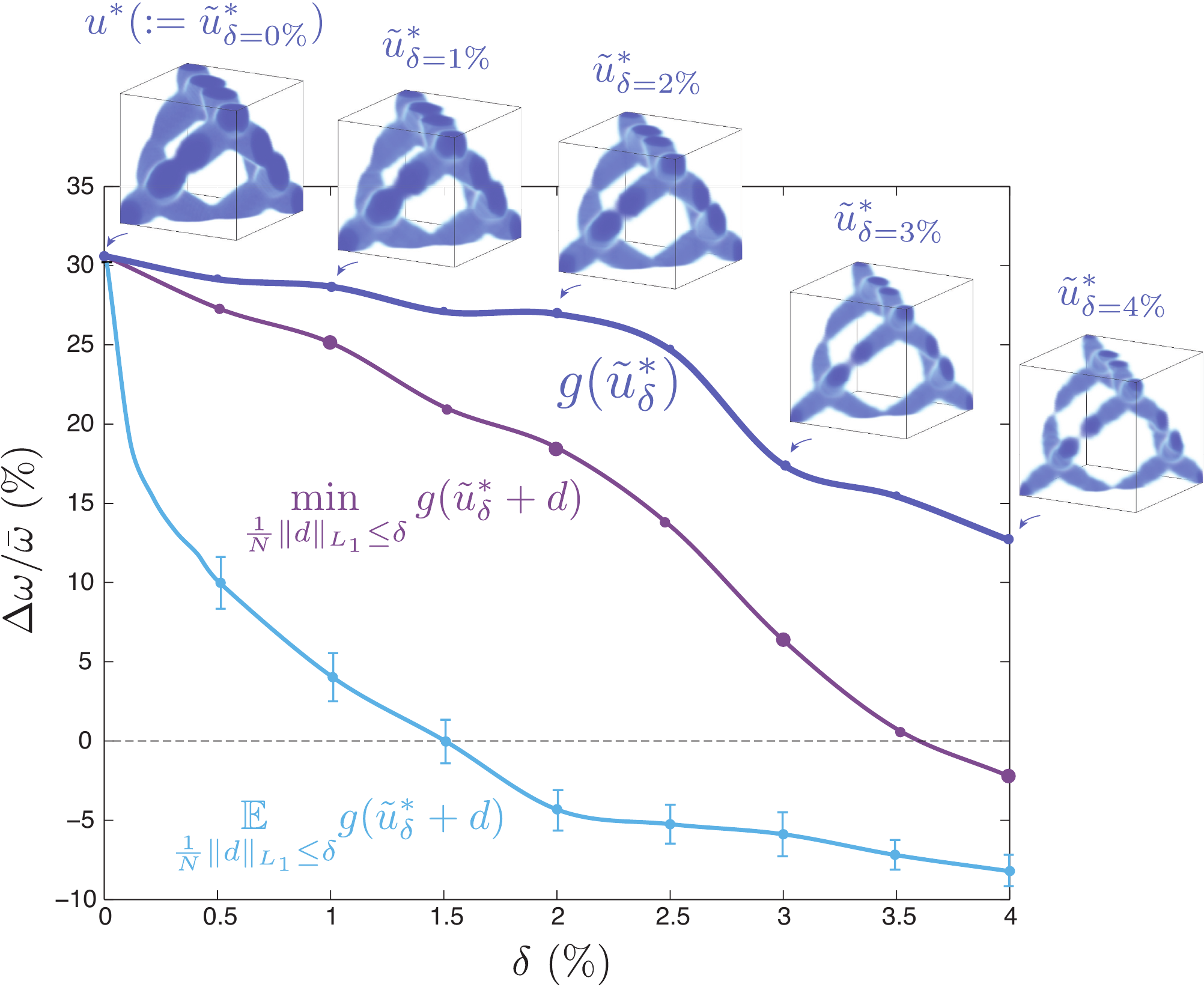}
\caption{Robust optimal designs for Diamond2, where $\tilde{u}^*_\delta$ solves the FA problem (\ref{opt_fa}) with uncertainty $\delta$ in $u$, and $\tilde{u}^*_0$ is the nominal optimum.  Top curve: gap size $g(\tilde{u}^*_\delta)$ of FA structure versus $\delta$, showing tradeoff between gap size and robustness to greater uncertainties.   Middle curve: worst-case gap $\min_d g(\tilde{u}^*_\delta + d)$ of the FA structure plus errors, which degrades the gap size.  Bottom curve: much greater degradation of gap size is found if we add uncertainty to the nominal (non-robust) structure $\tilde{u}^*_0$.  (In this case, we show the mean gap size for random perturbations $d$, with the standard deviation shown as error bars; this is an upper bound on the worst-case degradation.) }
\label{fig_gap_delta}
\end{figure}

Precisely this behavior is shown in \figref{fig_gap_delta}.  The top-most curve is the gap size of the FA-optimized structure as a function of $\delta$.   In particular, we let $\tilde{u}^*_\delta$ denote the geometry parameters found in solution to the FA optimization problem (\ref{opt_fa}) for a given $\delta$ (so that $\tilde{u}^*_0$ is the nominal optimum).  The top curve in  \figref{fig_gap_delta} is $g(\tilde{u}^*_\delta)$, the fractional gap from the the $\tilde{u}^*_\delta$, and as expected the gap decreases with $\delta$, decreasing especially rapidly for $\delta \gtrsim 2.5\%$.   This represents a tradeoff between robustness to increasing uncertainties versus performance (gap size). The insets show the corresponding structures: similar to our results in two dimensions~\cite{men2014fabrication}, the optimal dielectric veins are thinner in the presence of systematic uncertainties.

After the FA-optimized structure is found, we verified that the structure $\tilde{u}^*_\delta$ was much more robust to systematic errors than the non-robust structure $\tilde{u}^*_0$. This is shown in the bottom two curves in \figref{fig_gap_delta}.   The middle curve shows $\min_d g(\tilde{u}^*_\delta + d)$: the worst-case gap in response to systematic errors $d$ (of mean size $\delta$) in the robust structure $\tilde{u}^*_\delta$. Naturally this is worse than the performance of $\tilde{u}^*_\delta$ without errors, and the gap vanishes for $\delta > 3.5\%$. The bottom curve shows a similar but much larger degradation of the nominally optimal structure $\tilde{u}^*_0$ in response to similar perturbations, and in this case the gap typically vanishes for $\delta > 1.5\%$.  For the nominal structure, for computational simplicity we only computed the mean (expected) gap size in response to randomly chosen perturbations $d$ (in 30 trials); since the mean gap is obviously an upper bound on the worst-case gap, the degradation of the mean gap is sufficient to show that this structure is much worse than the FA structure.

\section{Conclusion}
\label{sec_conc}

Our results show that full 3D topology optimization of photonic band gaps is feasible, but suggest that little room remains for improving upon existing band-gap sizes even when fabrication constraints are removed. Of course, it is impossible to completely rule out the possibility that a much larger gap, or a much lower index contrast, is possible, for two reasons.  First, the nonconvexity of the optimization problem makes it difficult to prove that one has obtained the global optimum in such a large parameter space, although the fact that we obtain only a limited number of local optima from a large number of random starting points suggests that a global optimum has been found. Second, our optimization only searches one symmetry group and one pair of bands at a time, and future work could continue searching more groups and band pairs---our method of imposing a symmetry group, via projections, makes this particularly easy (as illustrated by the variety of space groups we were able to explore in this paper, in contrast to previous topology-optimization work). In principle, one could extend our approach in order to avoid imposing the symmetry group, by making the lattice vectors degrees of freedom and omitting the symmetry projection of the grid. This would require a much more expensive calculation because it would entail searching the entire Brillouin zone (or equivalently the entire unit cell in $\vec{k}$ space), but this may be feasible with a supercomputer-scale calculation or by a more clever method that performs an inner optimization over $\vec{k}$. However, we conjecture that such a {\it tour de force} would merely confirm that the high-symmetry structures are optimal, because the requirement of an omnidirectional gap tends to favor high-symmetry structures in order to have the same gap in multiple directions. (Nevertheless, the existence of large gaps in structures such as the single gyroid, which have only moderately large symmetry groups, lends some encouragement to a more thorough search.)

Regardless of the need for more band-gap designs, which is likely to be mainly driven by the discovery of new fabrication methods, the feasibility of 3D gap optimization offers the prospect of 3D topology optimization for many other dispersion-engineering problems. For example, our methods could be easily adapted to optimize superprism effects \cite{kosaka1998superprism,lin1996highly,kosaka1999superprism,wu2002superprism,luo2004superprism,serbin2005superprism}, supercollimation \cite{kosaka1999self,witzens2002self,wu2003beam,prather2004dispersion,shin2005conditions,lu2006experimental,rakich2006achieving}, dispersion compensation \cite{birks1999dispersion,shen2003design,ni2004dual,zsigri2004novel}, phase matching for nonlinear optics \cite{fiore1998phase,berger1998nonlinear,saltiel2000phase}, negative refraction for imaging \cite{notomi2000theory,luo2002pho,parimi2003photonic} or other dispersion constraints for various mode-coupling and mode-conversion problems \cite{prather2006photonic}. The SDP approach is essentially the same regardless of whether one is maximizing a band gap or minimizing the difference between $\omega(\vec{k})$ and some desired dispersion shape, and the need for robustness to fabrication variation arises in many such applications.

\section*{Acknowledgments}
H. Men, R. M. Freund, and J. Peraire were supported in part by the AFOSR Grant No. FA9550-11-1-0141, and the MIT--Chile--Pontificia Universidad Catlica de Chile Seed Fund. S.G. Johnson and K.Y.K. Lee were supported in part by the AFOSR MURI for Complex and Robust On-chip Nanophotonics (grant number FA9550-09-1-0704) and by the Institute for Soldier Nanotechnologies under Contract W911NF-07-D0004. 

\end{document}